\documentclass[apj]{emulateapj}

\submitted{ApJ, \textbf{698}, 1467--1484 (2009) [Accepted 14/04/2009]}

\newcommand{\be}{\begin{equation}}
\newcommand{\ee}{\end{equation}}
\newcommand{\bea}{\begin{eqnarray}}
\newcommand{\eea}{\end{eqnarray}}
\newcommand{\ha}{HI}
\newcommand{\hm}{H$_2$}
\newcommand{\avg}[1]{\langle#1\rangle}
\newcommand{\hfraction}{\beta}

\newcommand{\x}{x}
\newcommand{\xc}{X_{\rm c}}
\newcommand{\xv}{X_{\rm v}}
\newcommand{\correction}{\zeta}
\newcommand{\h}{h}
\newcommand{\freq}{\nu}
\newcommand{\BT}{\mathcal{B}}
\newcommand{\ggs}{f_{\sigma}}
\newcommand{\ggsc}{f_{\sigma}^0}
\newcommand{\gs}{\avg{\ggs}}

\newcommand{\f}{R_{\rm mol}}
\newcommand{\fc}{\f^{\rm c}}
\newcommand{\fg}{\f^{\rm galaxy}}
\newcommand{\rh}{r_{\rm s}}
\newcommand{\rdstars}{\tilde{r}_{\rm disk}}
\newcommand{\rdgas}{r_{\rm disk}}
\newcommand{\rd}{r_{\rm disk}}
\newcommand{\rb}{r_{\rm bulge}}
\newcommand{\rbp}{r_{\rm Plummer}}
\newcommand{\rha}{r_{\rm HI}}
\newcommand{\rhm}{r_{\rm H_2}}
\newcommand{\rhamax}{r_{\rm HI}^{\rm max}}
\newcommand{\rhmmax}{r_{\rm H_2}^{\rm max}}
\newcommand{\rvir}{r_{\rm vir}}

\newcommand{\ch}{c_{\rm halo}}
\newcommand{\cd}{c_{\rm disk}}
\newcommand{\cb}{c_{\rm bulge}}
\newcommand{\vc}{V_{\rm c}}
\newcommand{\vo}{V_{\rm obs}}
\newcommand{\vch}{V_{\rm c}^{\rm halo}}
\newcommand{\vcd}{V_{\rm c}^{\rm disk}}
\newcommand{\vcb}{V_{\rm c}^{\rm bulge}}
\newcommand{\vg}{\sigma_{\rm gas}}
\newcommand{\vsz}{\sigma_{\rm stars,z}}
\newcommand{\vgz}{\sigma_{\rm gas}}

\newcommand{\whafifty}{w_{\rm HI}^{50}}
\newcommand{\wcofifty}{w_{\rm CO}^{50}}
\newcommand{\whatwenty}{w_{\rm HI}^{20}}
\newcommand{\wcotwenty}{w_{\rm CO}^{20}}
\newcommand{\whapeak}{w_{\rm HI}^{\rm peak}}
\newcommand{\wcopeak}{w_{\rm CO}^{\rm peak}}
\newcommand{\mass}{M}
\newcommand{\msun}{{\rm M}_{\odot}}
\newcommand{\mhydro}{\mass_{\rm H}}
\newcommand{\mh}{\mass_{\rm halo}}
\newcommand{\md}{\mass^{\rm disk}}
\newcommand{\mb}{\mass^{\rm bulge}}
\newcommand{\mg}{\mass_{\rm gas}}
\newcommand{\mha}{\mass_{\rm HI}}
\newcommand{\mhm}{\mass_{{\rm H}_2}}
\newcommand{\mhe}{\mass_{\rm He}}
\newcommand{\ms}{\mass_{\rm stars}}
\newcommand{\msdisk}{\mass_{\rm stars}^{\rm disk}}
\newcommand{\msbulge}{\mass_{\rm stars}^{\rm bulge}}
\newcommand{\mvir}{\mass_{\rm vir}}

\newcommand{\mz}{\mass_{\rm Z}}
\newcommand{\mx}{\mass_{\rm x}}
\newcommand{\mbary}{\mass_{\rm bary}}
\newcommand{\Sigmad}{\Sigma^{\rm disk}}
\newcommand{\Sigmab}{\Sigma^{\rm bulge}}
\newcommand{\Sigmag}{\Sigma_{\rm gas}}
\newcommand{\Sigmas}{\Sigma_{\rm stars}}
\newcommand{\Sigmasdisk}{\Sigma_{\rm stars}^{\rm disk}}
\newcommand{\Sigmasbulge}{\Sigma_{\rm stars}^{\rm bulge}}

\newcommand{\Sigmaha}{\Sigma_{\rm HI}}
\newcommand{\Sigmahm}{\Sigma_{\rm H_2}}
\newcommand{\Sigmanull}{\Sigma_0}
\newcommand{\Sigmastd}{\tilde{\Sigma}_{\rm H}}
\newcommand{\Sigmahamax}{\Sigma_{\rm HI}^{\rm max}}
\newcommand{\Sigmahmmax}{\Sigma_{\rm H_2}^{\rm max}}
\newcommand{\Omegaha}{\Omega_{\rm HI}}
\newcommand{\Omegahm}{\Omega_{{\rm H}_2}}
\newcommand{\Omegag}{\Omega_{\rm gas}}
\newcommand{\rhoc}{\rho_{\rm c}}
\newcommand{\rhob}{\rho_{\rm bulge}}

\newcommand{\rhoh}{\rho_{\rm halo}}
\newcommand{\rhox}{\rho_{\rm x}}
\newcommand{\phix}{\phi_{\rm x}}
\newcommand{\phiha}{\phi_{\rm HI}}
\newcommand{\potd}{\varphi_{\rm disk}}
\newcommand{\lha}{L_{\rm HI}}

\newcommand{\lhacenter}{\Psi_{\rm HI}^0}
\newcommand{\lhamax}{\Psi_{\rm HI}^{\rm max}}
\newcommand{\lco}{L_{\rm CO}}

\newcommand{\lcocenter}{\Psi_{\rm CO}^0}
\newcommand{\lcomax}{\Psi_{\rm CO}^{\rm max}}

\shorttitle{Cosmological Simulation of Neutral Hydrogen}
\shortauthors{Obreschkow et al.}

\begin{document}

\title{Simulation of the Cosmic Evolution of Atomic and Molecular Hydrogen in Galaxies}

\author{D. Obreschkow$^1$, D. Croton$^2$, G. De Lucia$^3$, S. Khochfar$^4$, and S. Rawlings$^1$}

\affil{\vspace{0.4cm}\begin{flushleft}\small{
\hspace{0.85cm}$^1$~Astrophysics, Department of Physics, University of Oxford, Keble Road, Oxford, OX1 3RH, UK\\
\hspace{0.85cm}$^2$~Centre for Astrophysics and Supercomputing, Swinburne University of Technology, Mail H39, PO Box 218,\\
\hspace{1.13cm}Hawthorn, Victoria 3122, Australia\\
\hspace{0.85cm}$^3$~Max--Planck--Institut f\"ur Astrophysik, Karl--Schwarzschild--Str.~1, D-85748 Garching, Germany\\
\hspace{0.85cm}$^4$~Max--Planck--Institut f\"ur Extraterrestrische Physik, Giessenbachstr., D-85748, Garching, Germany}
\end{flushleft}}

\begin{abstract}
We present a simulation of the cosmic evolution of the atomic and molecular phases of the cold hydrogen gas in about $3\cdot10^7$ galaxies, obtained by post-processing the virtual galaxy catalog produced by \citet{DeLucia2007} on the Millennium Simulation of cosmic structure \citep{Springel2005}. Our method uses a set of physical prescriptions to assign neutral atomic hydrogen (\ha) and molecular hydrogen (\hm) to galaxies, based on their total cold gas masses and a few additional galaxy properties. These prescriptions are specially designed for large cosmological simulations, where, given current computational limitations, individual galaxies can only be represented by simplistic model-objects with a few global properties. Our recipes allow us to (i) split total cold gas masses between \ha, \hm, and Helium, (ii) assign realistic sizes to both the \ha- and \hm-disks, and (iii) evaluate the corresponding velocity profiles and shapes of the characteristic radio emission lines. The results presented in this paper include the local \ha- and \hm-mass functions, the CO-luminosity function, the cold gas mass--diameter relation, and the Tully--Fisher relation (TFR), which all match recent observational data from the local Universe. We also present high-redshift predictions of cold gas diameters and the TFR, both of which appear to evolve markedly with redshift. \\
\end{abstract}


\section{Introduction}\label{introduction}

Observations of gas in galaxies play a vital role in many fields of astrophysics and cosmology. Detailed studies of atomic and molecular material now possible in the local Universe with radio and millimeter telescopes will, over the coming decades, be extended to high redshifts as new facilities come on line.

Firstly, hydrogen is the prime fuel for galaxies, when it condenses from the hot ionized halo onto the galactic disks. The fresh interstellar medium (ISM) thus acquired mainly consists of atomic hydrogen (\ha), but in particularly dense regions, called molecular clouds, it can further combine to molecular hydrogen (\hm). Only inside these clouds can new stars form. Mapping \ha~and \hm~in individual galaxies therefore represents a key tool for understanding their growth and evolution. Secondly, the characteristic \ha-radio line permits the measurement of the radial velocity and velocity dispersion of the ISM with great accuracy, thereby leading to solid conclusions about galaxy dynamics and matter density profiles. Thirdly, and particularly with regard to next-generation radio facilities, surveys of \ha~are also discussed as a powerful tool for investigating the large scale structure of the Universe out to high redshifts. While such large scale surveys are currently dominated by the optical and higher frequency bands [e.g.~Spitzer \citep{Fang2005}, SDSS \citep{Eisenstein2005}, DEEP2 \citep{Davis2003}, 2dFGRS \citep{Cole2005}, GALEX \citep{Milliard2007}, Chandra \citep{Gilli2003}], they may well be overtaken by future radio arrays, such as the Square Kilometre Array \citep[SKA,][]{Carilli2004b}. The latter features unprecedented sensitivity and survey speed characteristics regarding the \ha-line and could allow the construction of a three-dimensional map of $\sim10^9$ \ha-galaxies in just a few years survey time. The cosmic structure hence revealed, specifically the baryon acoustic oscillations (BAOs) manifest in the power spectrum, will, for example, constrain the equation of state of dark energy an order of magnitude better than possible nowadays \citep{Abdalla2005,Abdalla2008}. Fourthly, deep low frequency detections will presumably reveal \ha~in the intergalactic space of the dark ages \citep{Carilli2004} -- one of the ultimate jigsaw pieces concatenating the radiation dominated early Universe with the matter dominated star-forming Universe.

Typically, \ha- and \hm-observations are considered part of radio and millimeter astronomy, as they rely on the characteristic radio line of \ha~at a rest-frame frequency of $\freq=1.42$ GHz and several carbon monoxide (CO) radio lines, indirectly tracing \hm-regions, in the $\freq=10^2-10^3$ GHz band. Such line detections will soon undergo a revolution with the advent of new radio facilities such as the SKA and the Atacama Large Millimeter/submillimeter Array (ALMA). These observational advances regarding \ha~and \hm~premise equally powerful theoretical predictions for both the optimal design of the planned facilities and for the unbiased analysis of future detections.

This is the first paper in a series of papers aiming at predicting basic \ha- and \hm-properties in a large sample of evolving galaxies. Here, we introduce a suite of tools to assign \ha- and \hm-properties, such as masses, disk sizes, and velocity profiles, to simulated galaxies. These tools are subsequently applied to the $\sim3\cdot10^7$ simulated evolving galaxies in the galaxy-catalog produced by \citet{DeLucia2007} (hereafter the ``DeLucia-catalog'') for the Millennium Simulation of cosmic structure \citep{Springel2005}. In forthcoming publications, we will specifically investigate the cosmic evolution of \ha- and \hm-masses and -surface densities predicted by this simulation, and we will produce mock-observing cones, from which predictions for the SKA and the ALMA will be derived.

Section \ref{section_background} provides background information about the DeLucia-catalog. In particular, we highlight the hybrid simulation scheme that separates structure formation from galaxy evolution, and discuss the accuracy of the cold gas masses of the DeLucia-catalog. In section \ref{section_masses}, we derive an analytic model for the \hm/\ha-ratio in galaxies in order to split the hydrogen of the cold gas of the DeLucia-catalog between \ha~and \hm. We compare the resulting mass functions (MFs) and the CO-luminosity function (LF) with recent observations. Sections \ref{section_sizes} and \ref{section_velocities} explain our model to assign diameters and velocity profiles to \ha- and \hm-disks. The simulation results are compared to observations from the local Universe and high-redshift predictions are presented. In Section \ref{section_discussion}, we discuss some consistency aspects and limitations of the approaches taken in this paper. Section \ref{section_conclusion} concludes the paper with a brief summary and outlook.

\section{Background: simulated galaxy catalog}\label{section_background}

$N$-body simulations of cold dark matter (CDM) on supra-galactic scales proved to be a powerful tool to analyze the non-linear evolution of cosmic structure \citep[e.g.][]{Springel2006}. Starting with small primordial perturbations in an otherwise homogeneous part of a model-universe, such simulations can quantitatively reproduce the large-scale structures observed in the real Universe, such as galaxy clusters, filaments, and voids. These simulations further demonstrate that most dark matter aggregations, especially the self-bound haloes, grow hierarchically, that is through successive mergers of smaller progenitors. Hence, each halo at a given cosmic time can be ascribed a ``merger tree'' containing all its progenitors. One of the most prominent simulations is the ``Millennium run'' \citep{Springel2005}, which followed the evolution of $2160^3\approx10^{10}$ particles of mass $8.6\cdot10^8\,\msun/\h$ over a redshift range $z=127\rightarrow0$ in a cubic volume of $(500\,{\rm Mpc}/\h)^3$ with periodic boundary conditions. The dimensionless Hubble parameter $\h$, defined as $H_0=100\,h\rm\,km\,s^{-1}\,Mpc^{-1}$, was set equal to $\h=0.73$, and the other cosmological parameters were chosen as $\Omega_{\rm matter}=0.25$, $\Omega_{\rm baryon}=0.045$, $\Omega_\Lambda=0.75$, $\sigma_8=0.9$.

Despite impressive results, modern $N$-body simulations of CDM in comoving volumes of order $(500\,{\rm Mpc}/h)^3$ cannot simultaneously evolve the detailed substructure of individual galaxies. The reasons are computational limitations, which restrict both the mass-resolution and the degree to which baryonic and radiative physics can be implemented. Nevertheless, an efficient approximate solution for the cosmic evolution of galaxies can be achieved by using a hybrid model that separates CDM-dominated structure growth from more complex baryonic physics \citep{Kauffmann1999}. The idea is to first perform a purely gravitational large-scale $N$-body simulation of CDM and to reduce the evolving data cube to a set of halo merger trees. These dark matter merger trees are assumed independent of the baryonic and radiative physics taking place on smaller scales, but they constitute the mass skeleton for the formation and evolution of galaxies. As a second step, each merger tree is populated with a list of galaxies, which are represented by simplistic model-objects with a few global properties (stellar mass, gas mass, Hubble type, star formation rate, etc.). The galaxies are formed and evolved according to a set of physical prescriptions, often of a ``semi-analytic'' nature, meaning that galaxy properties evolve analytically unless a merger occurs. This hybrid approach tremendously reduces the computational requirements compared to hydro-gravitational $N$-body simulations of each galaxy.

\cite{Croton2006} were the first to apply this hybrid scheme to the Millennium Simulation, thus producing a catalog with $\sim1.1\cdot10^9$ galaxies in 64 time steps, corresponding to about $\sim3\cdot10^7$ evolving galaxies. This catalog was further improved by \citet{DeLucia2007}, giving rise to the DeLucia-catalog used in this paper. The underlying semi-analytic prescriptions to form and evolve galaxies account for the most important mechanisms known today. In brief, the hot gas associated with the parent halo is converted into galactic cold gas according to a cooling rate that scales with redshift and depth of the halo potential. Stars form at a rate proportional to the excess of the cold gas density above a critical density, below which star formation is suppressed. In return, supernovae reheat some fraction of the cold gas, and, if the energy injected by supernovae is large enough, their material can escape from the galaxy and later be reincorporated into the hot gas. In addition, when galaxies become massive enough, cooling gas can be reheated via feedback from active galaxy nuclei (AGNs) associated with continuous or merger-based black hole mass accretion. These basic mechanisms are completed with additional prescriptions regarding merger-related starbursts, morphology changes, metal enrichment, dust evolution and change of photometric properties \citep[see][]{Croton2006,DeLucia2007}. The free parameters in this model were adjusted such that the simulated galaxies at redshift $z=0$ fit the joint luminosity/color/morphology distribution of observed low-redshift galaxies \citep{Cole2001,Huang2003,Norberg2002}. A good first order accuracy of the model is suggested by its ability to reproduce the observed bulge-to-black hole mass relation \citep{Haering2004}, the Tully--Fisher relation \citep{Giovanelli1997}, and the cold gas metallicity as a function of stellar mass (\citealp{Tremonti2004}, see also Figs.~4 and 6 in \citealp{Croton2006}).

Some galaxies in the DeLucia-catalog have no corresponding halo in the Millennium Simulation. Such objects can form during a halo merger, where the resulting halo is entirely ascribed to the most massive progenitor galaxy. In the model, the other galaxies continue to exist as ``satellite galaxies'' without haloes. These galaxies are identified as ``type 2'' objects in the DeLucia-catalog. If the halo properties of a satellite galaxy are required, they must be extrapolated from the original halo of the galaxy or estimated from the baryonic properties of the galaxy.

We emphasize that the semi-analytic recipes of the DeLucia-catalog are simplistic and may require an extension or readjustment, when new observational data become available. In particular, recent observations \citep{Bigiel2008,Leroy2008} suggest that star formation laws based on a surface density threshold are suspect, especially in low surface density systems. Moreover, galaxies in the DeLucia-catalog with stellar masses $\ms$ below $4\cdot 10^9\,\msun$ typically sit at the centers of haloes with less than 100 particles, whose merging history could only be followed over a few discrete cosmic time steps. It is likely that the physical properties of these galaxies are not yet converged. \cite{Croton2006} noted that especially the morphology (colors and bulge mass) of galaxies with $\ms\lesssim4\cdot 10^9\,\msun$ is poorly resolved, since, according to the model, the bulge formation directly relies on the galaxies' merging history and disk instabilities. Nevertheless, the simulated cosmic space densities of stars in early-type and late-type galaxies at redshift $z=0$ and the cosmic star formation history are consistent with observations \citep[see figures and references in][]{Croton2006}. It is therefore probable that at least the more massive galaxies in the simulation are not significantly affected by the mass-resolution and the simplistic law for star formation.

In this paper, we post-process the DeLucia-catalog to estimate realistic \ha- and \hm-properties for each galaxy. Our prescriptions will make use of the cold gas masses given for each galaxy in the DeLucia-catalog, and hence it is crucial to verify these cold gas masses against current observations. Based on a new estimation of the \hm-MF, we have recently calculated the normalized density of cold neutral gas in the local Universe as $\Omegag^{\rm obs}=(4.4\pm0.8)\cdot10^{-4}\ \h^{-1}$, hence $\Omegag^{\rm obs}\approx6.0\cdot10^{-4}$ for $\h=0.73$ \citep{Obreschkow2009a}. This value was obtained by integrating the best fitting Schechter functions of the local \ha-MF and \hm-MF and therefore it includes an extrapolation towards masses below the respective detection limits of \ha~and \hm. The simulated local cold gas density $\Omegag^{\rm sim}$ of the DeLucia-catalog, obtained from the sum of the cold gas masses of all galaxies at redshift $z=0$, exceeds the observed value by a factor $\correction=\Omegag^{\rm sim}/\Omegag^{\rm obs}=1.45$, as shown in Table \ref{tab_gasinsims}. For comparison, this table also lists the cold gas densities of other galaxy catalogs produced for the Millennium Simulation, using different semi-analytic recipes \citep{Bertone2007} and different schemes for the construction of dark matter merger trees \citep{Bower2006}.

\begin{table}[h]
  \centering
\begin{tabular}{|l|c|c|}
\hline
 Catalog              & $\Omegag^{\rm sim}$ & $\Omegag^{\rm sim}/\Omegag^{\rm obs}$ \\
\hline
 \citet{DeLucia2007}  & $8.7\cdot10^{-4}$  & 1.45 \\
 \citet{Bower2006}    & $14.8\cdot10^{-4}$ & 2.45 \\
 \citet{Bertone2007}  & $9.0\cdot10^{-4}$  & 1.50 \\
\hline
\end{tabular}
   \caption{Normalized cold gas densities at $z=0$ of three different semi-analytic galaxy simulations applied to the Millennium Simulation of cosmic structure. The rightmost column shows the multiplicative offset from the observed value as determined by \cite{Obreschkow2009a}.}
   \label{tab_gasinsims}
\end{table}

There are plausible reasons for the excess of cold gas in the DeLucia-catalog compared to observations. Most importantly, the semi-analytic recipes only distinguish between two gas phases: the hot ($T\approx10^6-10^7$~K) and ionized material located in the halo of the galaxy or group of galaxies, and the cold ($T\approx10^2-10^3$~K) gas in galactic disks. However, recent observations have clearly revealed that some hydrogen in the disk of the Milky Way is warm ($T\approx 10^4$~K) and ionized, too. For example, \citet{Reynolds2004} analyzed faint optical emission lines from hydrogen, helium, and trace atoms, leading to the conclusion that about $1/3$ of all the hydrogen gas in the Local Interstellar Cloud (LIC) is ionized. If this were true for all the gas in disk galaxies, one would expect a correction factor around 1.5 between simulated disk gas and cold neutral gas. Justified by this considerations, we decided to divide all the cold gas masses in the DeLucia-catalog $\mg^{\rm DeLucia}$ by the constant $\correction=1.45$ in order to obtain more realistic estimates,
\be\label{eqcorrection}
  \mg\equiv\correction^{-1}\mg^{\rm DeLucia}.
\ee

\section{Gas masses and mass functions}\label{section_masses}

In this section we establish a physical prescription to subdivide the cold and neutral hydrogen mass $\mhydro=\mha+\mhm$ of a galaxy into its atomic (\ha) and molecular (\hm) component based on the observed and theoretically confirmed relation between local gas pressure and local molecular fraction \citep{Elmegreen1993,Blitz2006,Leroy2008,Krumholz2009}. This prescription shall be applied to the DeLucia-catalog. The resulting simulated \ha- and \hm-mass functions (MFs) and the related CO-luminosity function (LF) will be compared to observations in the local Universe.

\subsection{Prescription for subdividing cold gas}\label{section_subdivision}

Before addressing the sub-composition of cold hydrogen, we note that the total cold hydrogen mass $\mhydro$ can be inferred from the total cold gas mass $\mg$ by a constant factor $\mhydro=0.74\,\mg$, which corresponds to the universal abundance of hydrogen \citep[e.g.][]{Arnett1996} that changes insignificantly with cosmic time. The remaining gas is composed of helium (He) and a minor fraction of heavier elements, collectively referred to as metals (Z). The DeLucia-catalog gives an estimate for the metal mass in cold gas $\mz$, and hence we shall compute the masses of cold hydrogen and He as
\be
\begin{array}{rcl}
  \mhydro  & = & (\mg-\mz)\cdot\hfraction, \label{splitcoldgas} \\
  \mhe     & = & (\mg-\mz)\cdot(1-\hfraction),
\end{array}
\ee
where the hydrogen fraction $\hfraction=0.75$ is chosen slightly above 0.74 to account for the subtraction of the 1--2\% metals in Eqs.~(\ref{splitcoldgas}).

The subdivision of the cold hydrogen mass $\mhydro=\mha+\mhm$ depends on the galaxy and evolves with cosmic time. We shall tackle this complexity using the variable \hm/\ha-ratio $\fg\equiv\mhm/\mha$, hence
\be
\begin{array}{rcl}
  \mha & = & \mhydro\cdot(1+\fg)^{-1}, \label{splithydrogen} \\
  \mhm & = & \mhydro\cdot(1+{\fg}^{-1})^{-1}.
\end{array}
\ee

Detailed observations of \ha~and CO in nearby regular spiral galaxies revealed that virtually all cold gas of these galaxies resides in flat, often approximately axially symmetric, disks \citep[e.g.][]{Walter2008,Leroy2008}. CO-maps recently obtained for five nearby elliptical galaxies \citep{Young2002} show that even these galaxies, who carry most of their stars are in a spheroid, have most of their cold gas in a disk. There is also empirical evidence, that most cold gas in high-redshift galaxies resides in disks \citep[e.g.][]{Tacconi2006}. Based on these findings, we assume that galaxies generally carry their cold atomic and molecular cold gas in flat disks with axially symmetric surface density profiles $\Sigmaha(r)$ and $\Sigmahm(r)$, where $r$ denotes the galactocentric radius in the plane of the disk. Using these functions, $\fg$ can be expressed as
\be\label{eqfgexact}
  \fg = \frac{2\pi\int_0^\infty{\rm d}r\,r\,\Sigmahm(r)}{2\pi\int_0^\infty{\rm d}r\,r\,\Sigmaha(r)}.
\ee

To solve Eq.~(\ref{eqfgexact}), we shall now derive an analytic model for $\Sigmaha(r)$ and $\Sigmahm(r)$. To this end, we analyzed the observed density profiles $\Sigmaha(r)$ and $\Sigmahm(r)$ presented by \cite{Leroy2008} for 12 nearby spiral galaxies of The HI Nearby Galaxy Survey (THINGS)\footnote{In total \cite{Leroy2008} analyzed 23 galaxies. Here, we only use the 12 galaxies, for which radial density profiles are provided for both \ha~and \hm~(based on CO(2--1) or CO(1--0) measurements), and we subtract the Helium-fraction included by \cite{Leroy2008}.}. In general, the surface density of the total hydrogen component (\ha+\hm) is well fitted by a single exponential profile,
\be
  \Sigmaha(r)+\Sigmahm(r) = \Sigmastd\exp(-r/\rdgas), \label{eqsig1}
\ee
where $\rdgas$ is a scale length and $\Sigmastd\equiv\mhydro/(2\pi\rdgas^2)$ is a normalization factor, which can be interpreted the maximal surface density of the cold hydrogen disk.

Eq.~(\ref{eqsig1}) can be solved for $\Sigmaha(r)$ and $\Sigmahm(r)$, if we know the local \hm/\ha-ratio in the disk, i.e.~the radial function $\f(r)\equiv\Sigmahm(r)/\Sigmaha(r)$. Following the theoretical prediction that $\f(r)$ scales as some power of the gas pressure \citep{Elmegreen1993}, \citet{Blitz2006} presented compelling observational evidence for this power-law based on 14 nearby spiral galaxies of various types. Perhaps the most complete empirical study of $\f(r)$ today has recently been published by \cite{Leroy2008}, who analyzed the correlations between $\f(r)$ and various disk properties in 23 galaxies of the THINGS catalog. This study confirmed the power-law relation between $\f(r)$ and pressure. On theoretical grounds, \cite{Krumholz2009} argued that $\f(r)$ is most fundamentally driven by density rather than pressure. However, by virtue of the thermodynamic relation between pressure and density, it is, in the context of this paper, irrelevant which quantity is considered, and the density-law for $\f(r)$ by \cite{Krumholz2009} is indeed consistent with the pressure-laws by \citet{Blitz2006} and \citet{Leroy2008}. Here, we shall apply the pressure-law
\be\label{eqblitz}
  \f(r) = [P(r)/P_\ast]^\alpha,
\ee
where $P(r)$ is the kinematic midplane pressure outside molecular clouds, and $P_\ast=2.35\cdot10^{-13}\ {\rm Pa}$ and $\alpha=0.8$ are empirical values adopted from \cite{Leroy2008}.

\cite{Elmegreen1989} showed that the equations of hydrostatic equilibrium for an infinite thin disk with gas and stars exhibit a simple approximate solution for the macroscopic kinematic midplane-pressure $P(r)$ of the ISM,
\be\label{eqpr}
  P(r) = \frac{\pi}{2}\,G\,\Sigmag(r)\Big(\Sigmag(r)+\ggs(r)\,\Sigmasdisk(r)\Big),
\ee
where $G$ is the gravitational constant, $\Sigmag(r)$ is the surface density of the total cold gas component (\ha$+$\hm$+$He$+$metals), $\Sigmasdisk(r)$ is the surface density of stars in the disk (thus excluding the bulge stars of early-type spiral galaxies and elliptical galaxies), and $\ggs(r)\equiv\vgz/\vsz$ is the ratio between the vertical velocity dispersions of gas and stars. The impact of supernovae and other small-scale effects on the gas pressure are implicitly included in Eq.~(\ref{eqpr}) via the velocity dispersion $\vgz$. For $\Sigmasdisk=0$, Eq.~(\ref{eqpr}) reduces to $P(r)=0.5\,\pi\,G\,\Sigmag(r)^2$, which is sometimes used as an approximation for the ISM pressure in gas-rich galaxies \cite[e.g.][]{Crosthwaite2007}.

To simplify Eq.~(\ref{eqpr}), we note that $\Sigmag(r)$ can be expressed as $\Sigmag(r)=\mg/(2\pi\,\rdgas^2)\exp(-r/\rdgas)$, which is identical to Eq.~(\ref{eqsig1}) up to the constant factor correcting for helium and metals. To find a similar expression for $\Sigmasdisk(r)$, we analyzed the stellar surface densities $\Sigmas(r)$ of the 12 THINGS galaxies mentioned before. In agreement with many other studies \citep[e.g.][]{Courteau1996}, we found that $\Sigmas(r)$ is generally well approximated by a double exponential profile, i.e.~the sum of an exponential profile $\Sigmasbulge$ for the bulge and an exponential profile $\Sigmasdisk$ for the disk. On average, the scale length of the stellar disk $\rdstars$ is 30\%--50\% smaller than the gas scale length $\rd$, which traces the fact that stars form in the more central \hm-dominated parts of galaxies. Indeed, several observational studies revealed that the stellar scale length is nearly identical to that of molecular gas \cite[e.g.][]{Young1995,Regan2001,Leroy2008}, and hence smaller than the scale length of \ha~or the scale length $\rdgas$ of the total cold gas component. For simplicity, we shall here assume $\rdgas=2\,\rdstars$ for all galaxies, such that $\Sigmasdisk(r)=4\,\ms/(2\pi\,\rdgas^2)\exp(-2\,r/\rdgas)$. Finally, we approximate the dispersion ratio $\ggs(r)$ as $\ggs(r)=\ggsc\,\exp(r/\rd)$, where $\ggsc$ is a constant. This approximation is motivated by empirical evidence that the gas dispersion $\vgz$ remains approximately constant across galactic discs \citep[e.g.][]{Dickey1990,Boulanger1992,Leroy2008}, combined with theoretical and observational studies showing that the stellar velocity dispersion $\vsz$ decreases approximately exponentially with a scale length twice that of the stellar surface density \citep[e.g.][]{Bottema1993}. Within those approximations Eq.~(\ref{eqpr}) reduces to
\be\label{eqpr2}
  P(r)\approx\frac{G\,\mg}{8\pi\,\rd^4}\Big(\!\mg\!+\!\gs\msdisk\!\Big)\exp(-2\,r/\rd),
\ee
where $\gs\equiv4\ggsc$ is a constant, which can be interpreted as the average value of $\ggs(r)$ weighted by the stellar surface density, since $\int 2\pi\,r\,\Sigmasdisk(r)\ggs(r)/\msdisk=4\ggsc$.

Substituting $P(r)$ in Eq.~(\ref{eqblitz}) for Eq.~(\ref{eqpr2}), we obtain
\be\label{eqfr}
  \frac{\Sigmahm(r)}{\Sigmaha(r)} \equiv \f(r) = \fc\,\exp(-1.6\,r/\rd)
\ee
with
\be\label{eqfc}
  \fc = \left[K\,\rd^{-4}\,\mg\,\Big(\mg+\gs\,\msdisk\Big)\right]^{0.8},
\ee
where $K\equiv G/(8\pi\,P_\ast)=11.3\rm\,m^4\,kg^{-2}$. Eq.~(\ref{eqfr}) reveals that the \hm/\ha-ratio $\f(r)$ is described by an exponential profile with scale length $\rd/1.6$. It should be emphasized that the central value $\fc$ does not necessarily correspond to the \hm/\ha-ratio measured at the center of real galaxies due to an additional \hm-enrichment caused by the central stellar bulge. However, $\fc$ represents the extrapolated cental \hm/\ha-ratio of the exponential profile, which approximates $\f(r)$ in the outer, disk-dominated galaxy parts.

\begin{figure}[h]
  \includegraphics[width=\columnwidth]{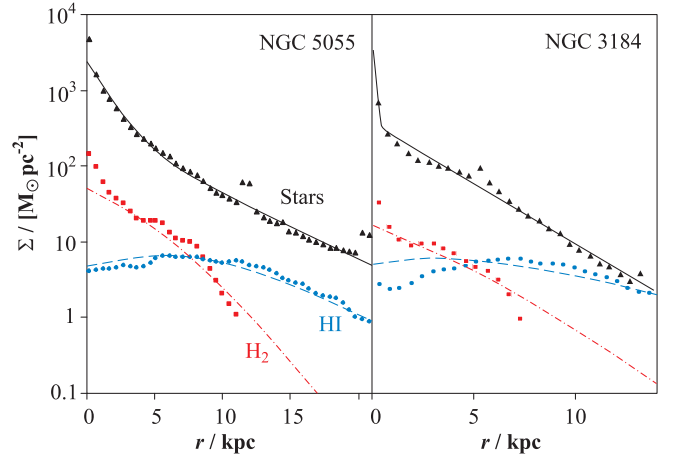}
  \caption{Column density profiles of two nearby spiral galaxies. Filled triangles, circles, and squares respectively represent the measured column density profiles of stars, \ha, and \hm~\citep{Leroy2008}. Solid lines show the best-fitting double-exponential functions for the stellar densities. Dashed lines and dash-dotted lines represent the predictions of our pressure-based model given in Eqs.~(\ref{eqsigmaha}, \ref{eqsigmahm}).}
  \label{fig_density_profiles}
\end{figure}

We can now solve Eqs.~(\ref{eqsig1}, \ref{eqfr}) for the atomic and molecular surface density profiles, i.e.
\bea
  \Sigmaha(r) & = & \frac{\Sigmastd\,\exp(-r/\rd)}{1+\fc\exp(-1.6\,r/\rd)}\ , \label{eqsigmaha} \\
  \Sigmahm(r) & = & \frac{\Sigmastd\,\fc\,\exp(-2.6\,r/\rd)}{1+\fc\exp(-1.6\,r/\rd)}\ , \label{eqsigmahm}
\eea
These model-profiles can be checked against the observed \ha- and \hm-density profiles of the nearby galaxies analyzed by \cite{Leroy2008}. In particular, we can test the limitations implied by our assumption that $\rdgas=2\,\rdstars$. To this end, we selected two observed regular spiral galaxies with  $\rdgas\approx2\,\rdstars$ (NGC 3184) and $\rdgas\approx\rdstars$ (NGC 5505). To evaluate Eqs.~(\ref{eqsigmaha}, \ref{eqsigmahm}) for those galaxies, we require the quantities $\msdisk$, $\mg$, $\rdgas$, and $\gs$. $\msdisk$, $\mg$, and $\rdgas$ were determined by fitting a single exponential profile to the total cold gas component and a double exponential profile (bulge and disk) to the stellar component, and $\gs$ was chosen as $\gs=0.4$, i.e.~the value given by \citet{Elmegreen1993} for nearby galaxies. As shown in Fig.~\ref{fig_density_profiles}, the resulting model-profiles $\Sigmaha(r)$ and $\Sigmahm(r)$ approximately match the empirical data. The fact that the fit is rather good for both galaxies demonstrates that the quality of the model-predictions does not sensibly depend on the goodness of the model-assumption $\rdgas\approx2\,\rdstars$. Similarly good fits are indeed found for most of the 12 THINGS-galaxies, for which \cite{Leroy2008} published radial \ha- and \hm-density profiles.

Eqs.~(\ref{eqsigmaha}, \ref{eqsigmahm}) can be solved for the maximal surface densities of \ha~and \hm. $\Sigmaha(r)$ exhibits its maximum at the radius $\rhamax=0.625\,\rdgas\,\ln(3/5\cdot\fc)$, as long as $\fc>5/3$. Galaxies in this category show an \ha-drop towards their center, such as observed in most galaxies in the THINGS catalog \citep{Walter2008}. By contrast, disk galaxies with $\fc\leq5/3$ have \ha-density profiles peaking at the center, $\rhamax=0$. Galaxies with such small values of $\fc$ have low gas densities by virtue of Eq.~(\ref{eqfc}), such as the irregular galaxies NGC 4214 and NGC 3077 \citep[see profiles in][]{Leroy2008}. $\Sigmahm(r)$ given in Eq.~(\ref{eqsigmahm}) always peaks at the disk center, $\rhmmax=0$.

The maximal values of $\Sigmaha(r)$ and $\Sigmahm(r)$, called $\Sigmahamax\equiv\Sigmaha(\rhamax)$ and $\Sigmahmmax\equiv\Sigmahm(\rhmmax)$, can be computed as
\bea
  \Sigmahamax/\Sigmastd &=& \left\{\begin{array}{ll}
    1/(1+\fc) & \rm{~if~}\fc\leq5/3\quad \\
    0.516\,{\fc}^{-5/8} & \rm{~if~}\fc>5/3
  \end{array}\right. \\
  \Sigmahmmax/\Sigmastd &=& \fc/(1+\fc).
\eea

The density profiles of Eqs.~(\ref{eqsigmaha}, \ref{eqsigmahm}) can be substituted into Eq.~(\ref{eqfgexact}). The exact solution of Eq.~(\ref{eqfgexact}) is quite unhandy, but $\fg$ only depends on $\fc$ and an excellent approximation, accurate to better than 5\% over the nine orders of magnitude $\fc=10^{-3}-10^6$ (covering the most extreme values at all redshifts), is given by
\be\label{eqfg}
  \fg = \big(3.44\,{\fc}^{-0.506}+4.82\,{\fc}^{-1.054}\big)^{-1}.
\ee

Eqs.~(\ref{eqfc}, \ref{eqfg}) constitute a physical prescription to estimate the \hm/\ha-ratio of any regular galaxy based on four global quantities: the disk stellar mass $\msdisk$, the cold gas mass $\mg$, the scale radius of the cold gas disk $\rdgas$, and the dispersion parameter $\gs$. In \cite{Obreschkow2009a}, we showed that the \hm/\ha-ratios inferred from this model are consistent with observations of nearby galaxies. Moreover, this model presumably extends to high redshifts, since it essentially relies on the fundamental relation between pressure and molecular fraction and on a few other physical assumptions with weak or absent dependence on cosmic epoch. However, a critical discussion of the limitations of this model is presented in Sections \ref{section_discussion_local} and \ref{section_discussion_evolution}.

\subsection{Application to the DeLucia-catalog}\label{section_appl}

We applied the model given in Eqs.~(\ref{eqfc}, \ref{eqfg}) together with Eqs.~(\ref{splitcoldgas}, \ref{splithydrogen}) to the DeLucia-catalog in order to assign \ha-, \hm-, and He-masses to the simulated galaxies. The quantities $\msdisk$ and $\mz$ used in these Eqs.~are directly contained in the DeLucia-catalog, and $\mg$ was inferred from the given cold gas masses via the correction of Eq.~(\ref{eqcorrection}).  The dispersion parameter $\gs$ is approximated by the constant $\gs=0.4$, consistent with the local observational data used by \cite{Elmegreen1989} (but see discussion in Section \ref{section_discussion_evolution}).

The remaining and most subtle ingredient for our prescription of Eqs.~(\ref{eqfc}, \ref{eqfg}) is the scale radius $\rd$. This radius can be estimated from the virial radius $\rvir$ of the parent halo, but their relation is intricate. Even modern $N$-body plus SPH simulations of galaxy formation cannot reproduce observed disk diameters \citep{Kaufmann2007}, and hence the more simplistic semi-analytic approaches are likely to require some empirical adjustment. \citet{Mo1998} studied the case of a flat exponential disk in an isothermal singular halo. When assuming that the disk's mass can be neglected for its rotation curve, they find
\be\label{eqrd}
  \rd = \frac{\lambda\cdot\xi}{\sqrt{2}}\,\rvir,
\ee
where $\lambda$ is the spin parameter of the halo and $\xi$ is the ratio between the specific angular momentum of the disk (angular momentum per unit mass) and the specific angular momentum of the halo. The model behind Eq.~(\ref{eqrd}) does not distinguish between different scale radii for stars and cold gas, but it assumes a single exponential disk in hydrostatic equilibrium without including the effects of star formation. It is therefore natural to identify $\rd$ in Eq.~(\ref{eqrd}) with the cold gas scale radius $\rdgas$ of Eqs.~(\ref{eqfc}--\ref{eqsigmahm}).

In the Millennium Simulation, $\rvir$ was calculated from the virial mass $\mvir$ using the relation
\be\label{eqmvir}
  \mvir = \frac{4\pi}{3}\rvir^3\cdot200~\rhoc(z)
\ee
where $\rhoc(z)$ is the critical density for closure $\rhoc(z)=3H^2(z)/(8\pi G)$ and $\mvir$ the virial mass of the halo. For central haloes, $\mvir$ was approximated as $M_{200}$, i.e.~the mass in the region with an average density equal to $200~\rhoc(z)$; and for sub-haloes, $\mvir$ was approximated as the total mass of the gravitationally bound simulation-particles.

The spin parameter $\lambda$ was calculated directly from the $N$-body Millennium Simulation according to the definition $\lambda\equiv J_{\rm halo}{¦E_{\rm halo}¦}^{1/2}G^{-1}{\mh}^{-5/2}$, where $J_{\rm halo}$ denotes the angular momentum of the halo, $E_{\rm halo}$ its energy, and $\mh$ its total mass. For the satellite galaxies in the DeLucia-catalog, i.e.~the ones without halo (see Section \ref{section_background}), the value of $\lambda\cdot\rvir$ was approximated as the respective value of the original galaxy halo before its disappearance.

The only missing parameter for the calculation of $\rd$ via Eq.~(\ref{eqrd}) is the angular momentum ratio $\xi$. It is of order unity for evolved disk-galaxies \citep[e.g.][]{Fall1980,Zavala2008}, but its exact value is uncertain because of the difficulty of measuring the spin of dark matter haloes and convergence issues in numerical simulations. For example, \citet{Kaufmann2007} showed that $N$-body plus SPH simulations with as many as $10^6$ particles per galaxy do not reach convergence in angular momentum, because of the difficulties to model the transport of angular momentum.

Here, we shall chose $\xi$, such that our simulation reproduces the empirical relation between the galaxy baryon mass $\mbary$ and the stellar scale radius $\rdstars\approx\rd/2$, measured in the local Universe. To ensure consistency with Section \ref{section_subdivision}, we use again the data from the THINGS-galaxies analyzed by \citep{Leroy2008}. Of the 23 galaxies in this sample, we reject the 6 irregular objects, since our model for \ha~and \hm~assumes regular galaxies. The remaining 17 galaxies cover all spiral types and include \ha-rich and \hm-rich galaxies. For each galaxy, we adopted the stellar scale radii $\rdstars$ and baryon masses $\mbary=\ms+\mha+\mhm$ directly from the data presented in \citep{Leroy2008}\footnote{Stellar masses $\ms$ rely on $3.6\,\mu\rm m$-maps \citep[SINGS,][]{Kennicutt2003}; \ha-masses use the $21\rm\,cm$-maps from THINGS \citep{Walter2008}, \hm-masses rely on CO-maps (CO(2--1) from HERACLES, \citealp{Leroy2008}; CO(1--0) from BIMA SONG, \citealp{Helfer2003}).}. The Hubble-types $T$, i.e.~the numerical stage indexes along the revised Hubble sequence of the RC2 system \citep{deVaucouleurs1976}, were drawn from the HyperLeda database \citep{Paturel2003}.

\begin{figure}[h]
  \includegraphics[width=\columnwidth]{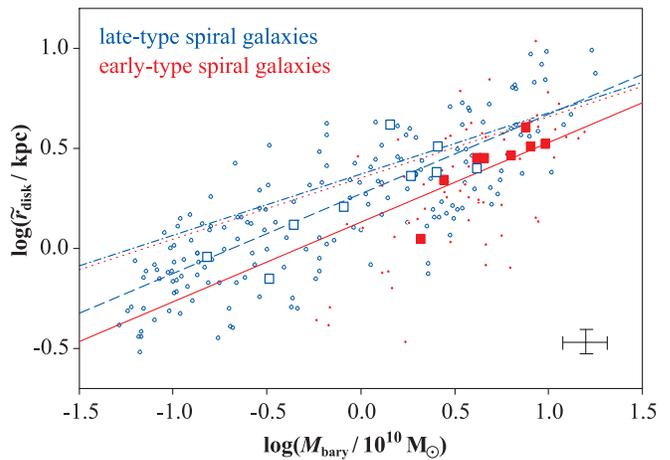}
  \caption{Relation between the baryon mass (stars+cold gas) and the stellar scale radius of disk galaxies. The filled and empty squares, respectively, represent observed early- and late-type spiral galaxies \citep{Leroy2008}. Typical 1-$\sigma$ error bars are shown in the bottom right corner. The empirical fit given in Eq.~(\ref{eqmrdfit}) is represented by a solid line for early-type spiral galaxies ($T=2$) and by a dashed line for late-type spiral galaxies ($T=10$). The dotted and dash-dotted lines represent the respective relations of the simulated galaxies in the DeLucia-catalog, if $\xi=1$. In order for the simulation to maximally align with the observations, $\xi$ must be chosen according to Eq.~(\ref{eqxi}). For the latter case, $10^2$ random early- and late-type spiral galaxies of the DeLucia-catalog at $z=0$ are represented by filled and empty dots, respectively.}
  \label{fig_mass_scaleradius}
\end{figure}

The resulting empirical relation between $\mbary$ and $\rdstars$ is displayed in Fig.~\ref{fig_mass_scaleradius}. The scatter of these data probably underestimates the true scatter caused by all galaxies, since we exclusively considered non-interacting regular spiral galaxies. The zero-point of the mean relation between $\mbary$ and $\rdstars$ depends on the morphological galaxy type, as is revealed by the distinction of early-type and late-type spiral galaxies in Fig.~\ref{fig_mass_scaleradius}. Given identical baryon masses, the stellar scale radii of early-type galaxies tend to be smaller than the radii of late-type galaxies. This trend was also detected in other data samples (e.g.~data from \citealp{Kregel2002} shown in \citealp{Obreschkow2009a}). Several reasons could explain this finding: (i) early-type galaxies have more massive stellar bulges, which present an additional central potential that contracts the disk; (ii) bulges often form from disk instabilities, occurring preferably in systems with relatively low angular momentum, and hence early-type galaxies are biased towards smaller angular momenta and smaller scale radii; (iii) larger bulges like those of lenticular and elliptical galaxies, often arise from galaxy mergers, which tend to reduce the specific angular momenta and scale radii.

To parameterize the dependence of the scale radius on the morphological galaxy type, the latter shall be quantified using the stellar mass fraction of the bulge, $\BT\equiv\msbulge/\ms$. In the observed sample, the values of $\BT$ can be approximately inferred from the Hubble type $T$. Here we shall use the relation
\be\label{eqtbs}
  \BT = (10-T)^2/256,
\ee
which approximately parameterizes the mean behavior of 146 moderately inclined barred and unbarred local spiral galaxies analyzed by \citet{Weinzirl2009}. Eq.~(\ref{eqtbs}) satisfies the boundary conditions $\BT=1$ for $T=-6$ (i.e.~pure spheriods) and $\BT=0$ for $T=10$ (pure disks).

We shall approximate the relation between $\mbary$ and $\rdstars$ as a power-law with an additional term for the observed secondary dependence on morphological type,
\be\label{eqmrdfit}
  \log\left(\frac{\rdstars}{\rm kpc}\right) = a_0+a_1\log\left(\frac{\mbary}{10^{10}\msun}\right)+a_2\,\BT,
\ee
where $a_0$, $a_1$, and $a_2$ are free parameters. The best fit to the empirical data in terms of a maximum-likelihood approach is given by the choice $a_0=0.3$, $a_1=0.4$, $a_2=-0.6$.

The fit of Eq.~(\ref{eqmrdfit}) is displayed in Fig.~\ref{fig_mass_scaleradius} for early-type spiral galaxies ($T=2\leftrightarrow\BT=0.25$, solid line) and for late-type spiral galaxies ($T=10\leftrightarrow\BT=0$, dashed line). For comparison, the mean power-law relations for the early- and late-type spiral galaxies in the DeLucia-catalog at $z=0$ are displayed as dotted and dash-dotted lines for the choice $\xi=1$. The simulated scale radii using $\xi=1$ are consistent with the observed ones for very massive galaxies ($\mbary\approx10^{11}\,\msun$), but less massive galaxies in the simulation turn out slightly too large if $\xi=1$. Furthermore, the morphological dependence of the simulation is too small compared to the observations. This can be corrected ad hoc by introducing a variable $\xi$ that depends on both $\mbary$ and $\BT$, i.e.
\be\label{eqxi}
  \log(\xi) = b_0+b_1\log\left(\frac{\mbary}{10^{10}\msun}\right)+b_2\,\BT,
\ee
where $b_0$, $b_1$, and $b_2$ are free parameters. The parameters minimizing the rms-deviation between the simulated galaxies and the empirical model of Eq.~(\ref{eqmrdfit}) are $b_0=-0.1$, $b_1=0.3$, $b_2=-0.6$. In addition, we chose a lower limit for $\xi$ equal to 0.5, in order to prevent unrealistically small scale radii.

We emphasize that Eq.~(\ref{eqxi}) is merely an empirical correction; this choice of $\xi$ should not be considered as an estimate of the true ratio between the specific angular momenta of the disk and the halo, but it also accounts for the imperfection of the simplistic halo model by \citet{Mo1998}, for missing physics in the semi-analytic modeling, and for possible systematic errors in the spin parameters $\lambda$ of the Millennium Simulation. The average value of Eq.~(\ref{eqxi}) over all galaxies in the DeLucia-catalog is $\avg{\xi}=0.7$ (with $\sigma=0.2$), which is approximately consistent with $\xi\approx1$ of modern high-resolution simulations of galaxy formation \citep[e.g.][]{Zavala2008}, even though the latter still suffer from issues with the transport of angular momentum as mentioned above.

Using Eqs.~(\ref{eqrd}, \ref{eqxi}) we estimated a scale radius $\rd$ for each galaxy in the DeLucia-catalog. A sample of $10^2$ simulated early- and late-type spiral galaxies at $z=0$ is shown in Fig.~\ref{fig_mass_scaleradius}. Given $\rd$ as well as $\ms$, $\mg$, and $\gs=0.4$, we then applied Eqs.~(\ref{splitcoldgas}, \ref{splithydrogen}, \ref{eqfc}, \ref{eqfg}) in order to subdivide the non-metallic cold gas mass $(\mg-\mz)$ of each galaxy into \ha, \hm, and He.

\subsection{Atomic and molecular mass functions}\label{subsection_mfs}

We shall now compare the \ha- and \hm-masses predicted by our model of Sections \ref{section_subdivision} and \ref{section_appl} to recent observations in the local Universe. From the viewpoint of the simulation, a fundamental output are the mass functions (MFs) of \ha~and \hm, while the available observational counterparts are the luminosity functions (LFs) of the \ha-emission line \citep{Zwaan2005} and the CO(1--0)-emission line \citep{Keres2003}. Therefore, either the simulated data or the observed data need a luminosity-to-mass (or vice versa) conversion to compare the two. Section \ref{subsection_mfs} focuses on the MFs, adopting the standard luminosity-to-mass conversion for \ha~used by \citet{Zwaan2005} and the CO-luminosity-to-\hm-mass conversion of \citet{Obreschkow2009a}. As a complementary approach, Section \ref{section_luminosities} will focus on the LFs, which will require a model for the conversion of simulated \hm-masses into CO-luminosities.

We define the MFs as $\phix(\mx)\equiv{\rm d}\rhox/{\rm d}\log\mx$, where $\rhox(\mx)$ is the space density (number per comoving volume) of galaxies containing a mass $\mx$ of the constituent x (\ha, \hm, He, etc.). Given a mass, such as $\mha$, for each galaxy in the DeLucia-catalog, the derivation of the corresponding MF only requires the counting of the number of sources per mass interval. We chose 60 mass intervals, logarithmically spaced between $10^8\,\msun$ and $10^{11}\,\msun$, giving about $\sim10^6$ galaxies per mass interval in the central mass range, while keeping the mass error relatively small ($\Delta\log(M)<0.05$). Since MFs combine units of mass and length, they generally depend on the Hubble constant $H_0$, or the dimensionless Hubble parameter $\h$, defined as $H_0=100\,\h$\,km\,s$^{-1}$\,Mpc$^{-1}$. Although MFs are often plotted in units making no assumption on $\h$ (e.g.~$\msun\,h^{-2}$ for the mass scale), this is impossible when observations are compared to cosmological simulations. The reason is that simulated masses in the Millennium Simulation scale to first order as $\h^{-1}$, whereas empirical masses, when determined from electromagnetic fluxes, are proportional to the square of the distance and hence scale as $\h^{-2}$. For all plots in this paper we shall therefore use $\h=0.73$, which corresponds to the value adopted by the Millennium Simulation (see Section \ref{section_background}).

\begin{figure}[h]
  \includegraphics[width=\columnwidth]{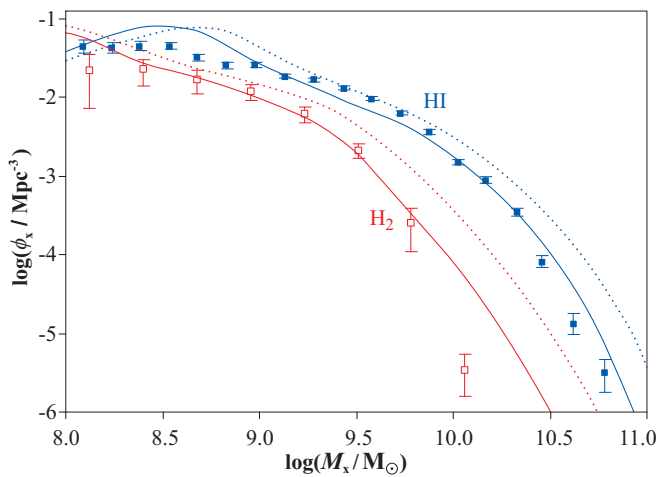}
  \caption{Simulated galaxy MFs for \ha~and \hm~with (solid lines) and without (dotted lines) the constant correction for all cold gas masses given in Eq.~(\ref{eqcorrection}). Filled and open squares with error bars represent the corresponding empirical MFs from \citet{Zwaan2005} and \citet{Obreschkow2009a}.}
  \label{fig_local_mf1}
\end{figure}

Fig.~\ref{fig_local_mf1} displays the \ha- and \hm-MF of our simulation (solid lines), as well as the corresponding empirical MFs for the local Universe (points with error bars). The empirical \ha-MF was obtained by \citet{Zwaan2005} based on 4315 galaxies of the HI-Parkes All Sky Survey (HIPASS) and the empirical \hm-MF was derived in \cite{Obreschkow2009a} from the CO-luminosity function (LF) presented by \citet{Keres2003}. Both empirical MFs approximately match the simulated data. We note, however, that the consistency between observation and simulation decreases if we skip the overall correction of the cold gas masses in the DeLucia-catalog by the constant factor $\correction$ (dotted lines), which, as argued in Section \ref{section_background} can be justified by a fraction of the disk gas being electronically excited or ionized.

Our simulation slightly over-predicts the observed number of the largest \ha- and \hm-masses, i.e.~the ones in the exponential tail of the MFs in Fig.~\ref{fig_local_mf1}. These tails contain the most massive systems, whose emergent luminosities are most likely to be biased by opacity and thermal effects. Including these effects would probably correct the space density of massive systems towards the simulated MFs. Additionally, we note that the presented empirical MFs neglect mass measurement errors, which might have an important effect on the slope of the exponential tails. Another difference between observations and simulation are the spurious bumps in the low mass range of the simulated MFs, i.e.~$\log(\mha)\approx8.5$ and $\log(\mhm)\approx8.0$, where the number density is about doubled compared to observations. This feature can also be seen in the optical b$_{\rm J}$-band LF shown by \citet{Croton2006} and stems from an imprecision in the number density of the smallest galaxies in the DeLucia-catalog, where the mass-resolution of the Millennium Simulation implies a poorly resolved merger history. The over-density of sources around this resolution limit roughly balances the mass of even smaller galaxies, i.e.~$\log(\mha/\msun)\ll8.0$, that are missing in the simulation.

The universal gas densities of the simulation, expressed relative to the critical density for closure, are $\Omegaha^{\rm sim}=3.4\cdot10^{-4}$ and $\Omegahm^{\rm sim}=1.1\cdot10^{-4}$, in good agreement with the observations $\Omegaha^{\rm obs}=(3.6\pm0.4)\cdot10^{-4}$ \citep{Zwaan2005} and $\Omegahm^{\rm obs}=(0.95\pm0.37)\cdot10^{-4}$ \citep{Obreschkow2009a}.

\begin{figure}[h]
  \includegraphics[width=\columnwidth]{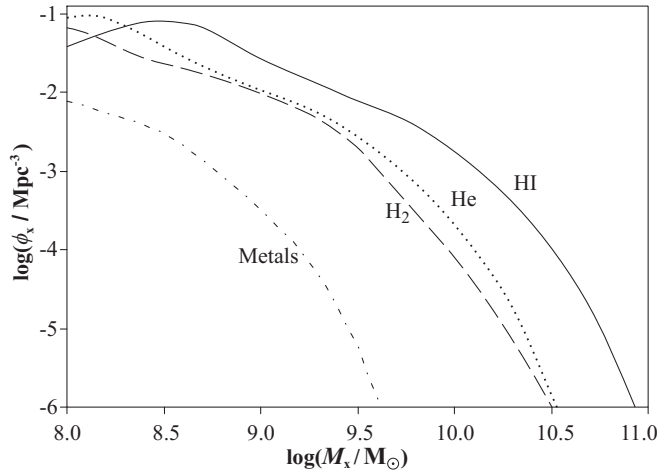}
  \caption{Simulated galaxy MFs for \ha~(solid line), \hm~(dashed line), cold He (dotted line), and cold gas metals (dash-dotted line). The \ha-MF and \hm-MF are identical to the solid lines in Fig.~\ref{fig_local_mf1}.}
  \label{fig_local_mf2}
\end{figure}

Fig.~\ref{fig_local_mf2} shows our simulated \ha-MF and \hm-MF together with the MF for the cold gas metals given in the original DeLucia-catalog and the MF for He as trivially derived using Eq.~(\ref{splitcoldgas}). This picture reveals that in the cold gas of the local Universe He is probably more abundant than \hm, but less abundant than \ha.

We shall now consider the \ha-masses in elliptical and spiral galaxies separately. This division is based on the Hubble type $T$, where we consider galaxies with $T<0$ as ``ellipticals'' and galaxies with $T\geq0$ as ``spirals'' -- a rough separation that neglects other types like irregular galaxies as well as various subclassifications. In the simulation, $T$ is computed from the bulge mass fraction according to Eq.~(\ref{eqtbs}).

\begin{figure}[h]
  \includegraphics[width=\columnwidth]{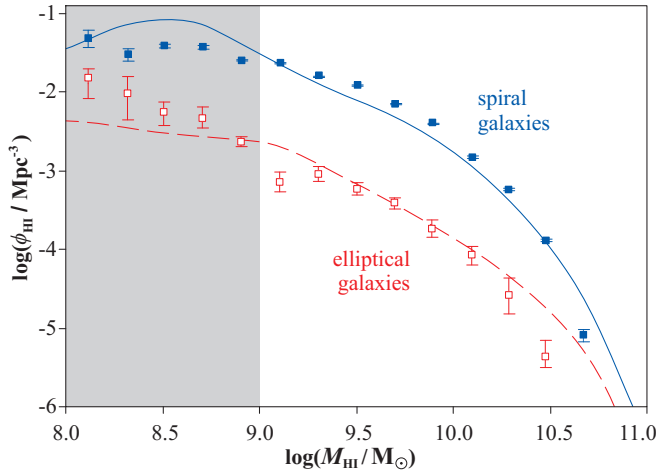}
  \caption{\ha-MFs for elliptical and spiral galaxies. The solid line and dashed line respectively represent the simulated result, where the galaxies have been divided in ellipticals and spirals according to their Hubble type estimated using Eq.~(\ref{eqtbs}). Filled and open dots with error bars represent the corresponding empirical \ha-MFs, which we derived from the HIPASS sample. The shaded zone represents the \ha-mass range $\mha\lesssim10^9\,\msun$, approximately corresponding to the simulated galaxies with poorly resolved morphologies.}
  \label{fig_local_mf_by_type}
\end{figure}

The simulated \ha-MFs of both elliptical and spiral galaxies are shown in Fig.~\ref{fig_local_mf_by_type}. In order to determine the observational counterparts, we split the HIPASS galaxy sample into elliptical and spiral galaxies according to the Hubble types provided in the HyperLeda reference database \citep{Paturel2003}. For both subsamples, the \ha-MF was evaluated using the $1/V_{\rm max}$ method \citep{Schmidt1968}, where $V_{\rm max}$ was estimated from the analytic completeness function for HIPASS, which characterizes the completeness of each source given its \ha-peak flux density $S_{\rm p}$ and integrated \ha-line flux $S_{\rm int}$ \citep{Zwaan2004}. In order to estimate the uncertainties of the MFs, we derived them for $10^4$ random half-sized subsets of the HIPASS sample -- a bootstrapping approach. The standard deviation of the $10^4$ values of $\log(\phiha)$ for each mass bin was divided by $\sqrt{2}$ to estimate the 1-$\sigma$ errors of $\log(\phiha)$ for the full sample.

Fig.~\ref{fig_local_mf_by_type} demonstrates that our simulation successfully reproduces the \ha-masses of both spiral and elliptical galaxies for HI-masses greater than $\sim10^9\,\msun$, although the nearly perfect match between simulation and observation may be somewhat coincidental due to the uncertainties of the Hubble types $T$ calculated via Eq.~(\ref{eqtbs}). For \ha-masses smaller than $10^9\,\msun$, the morphological separation seems to breakdown (shaded zone in Fig.~\ref{fig_local_mf_by_type}). Indeed the \ha-mass range $\mha\lesssim10^9\,\msun$ approximately corresponds to the stellar mass range $\ms\lesssim4\cdot10^9\,\msun$, for which morphology properties are poorly resolved (see Section \ref{section_background}).

\subsection{Observable HI- and CO-luminosities}\label{section_luminosities}

The characteristic radio line of \ha~stems from the hyperfine energy level splitting of the hydrogen atom and lies at 1.42 GHz rest-frame frequency. The velocity-integrated luminosity of this line $\lha$ can be calculated from the \ha-mass via
\be \label{eq_mha}
  \frac{\mha}{\msun} = 1.88\cdot10^4\cdot\frac{\lha}{\rm Jy\,km\,s^{-1}\,Mpc^2}.
\ee
Eq.~(\ref{eq_mha}) neglects \ha-self absorption effects, but this is likely to be a problem only for the largest disk galaxies observed edge-on \citep{Rao1995}. The strict proportionality between $\lha$ and $\mha$ assumed in Eq.~(\ref{eq_mha}) means that the \ha-LF is geometrically identical to the \ha-MF.

By contrast, the \hm-masses used for the empirical \hm-MFs in Figs.~\ref{fig_local_mf1} and \ref{fig_local_mf2} rely on measurements of the CO(1--0)-line, i.e.~the 115 GHz radio line stemming from the fundamental rotational relaxation of the most abundant CO-isotopomer $\rm ^{12}C^{16}O$. Here we only consider this line, but luminosities of other CO-lines can be estimated using approximate empirical line ratios \citep[e.g.][]{Braine1993b,Righi2008}.

The CO(1--0)-to-\hm~conversion generally depends on the galaxy and the cosmic epoch, and it is often represented by the dimensionless factor
\be \label{eqx}
  X \equiv \frac{N_{{\rm H}_2}/{\rm cm}^{-2}}{I_{\rm CO}/({\rm K\,km\,s}^{-1})}\cdot10^{-20},
\ee
where $N_{{\rm H}_2}$ is the column density of \hm-molecules and $I_{\rm CO}$ is the integrated CO(1--0)-line intensity per unit surface area defined via the surface brightness temperature in the Rayleigh-Jeans approximation. The definition of Eq.~(\ref{eqx}) implies the mass--luminosity relation \citep[e.g.~review by][]{Young1991}
\be \label{eq_mhm}
  \frac{\mhm}{\msun} = 313\cdot X\cdot\frac{\lco}{\rm Jy\,km\,s^{-1}\,Mpc^2},
\ee
where $\lco$ is the velocity-integrated luminosity of the CO(1--0) line.

As discussed in \cite{Obreschkow2009a}, the theoretical and observational determination of the $X$-factor is a subtle task with a long history. Most present-day studies assume a constant $X$-factor $\xc$, such as
\be\label{eqxconst}
  \xc = 2,
\ee
which is typical for spiral galaxies in the local Universe \citep{Leroy2008}. By contrast, \citet{Arimoto1996} and \citet{Boselli2002} suggested that $X$ is variable, $\xv$, and approximately inversely proportional to the metallicity $O/H$, i.e.~the ratio between the number of oxygen ions and hydrogen ions in the hot ISM. Using their data, we found that \citep{Obreschkow2009a}
\be\label{eqxoh}
  \log(\xv)=(-2.9\pm0.2)-(1.02\pm0.05)\log(O/H).
\ee
At first sight, the empirical negative dependence of $X$ on the metallicity seems to contradict the fact that $\rm ^{12}C^{16}O$ is optically thick for 115 GHz radiation. Indeed, the radiated luminosity should not depend on the density of metals, as long as the latter is high enough for the radiation to remain optically thick \citep{Kutner1985}. However, detailed theoretical investigations \citep[e.g.][]{Maloney1988} of the sizes and temperatures of molecular clumps were indeed able to explain, and in fact predict, the negative dependence of $X$ on metallicity.

Eq.~(\ref{eqxoh}) links the metallicity of the hot ISM to the $X$-factor of cold molecular clouds, and it is likely a consequence of a more fundamental relation between cold gas metallicity and $X$. To uncover such a relation, we assume that the $O/H$ metallicity of the cold ISM in local galaxies is approximated by $O/H$ of the hot ISM. Given an atomic mass of 16 for Oxygen, the fact that hydrogen makes up a fraction 0.74 of the total baryon mass, and assuming that Oxygen accounts for a fraction of 0.4 of the mass of all metals \citep[based on][]{Arnett1996,Kobulnicky1999}, Eq.~(\ref{eqxoh}) translates to
\be\label{eqxsim}
  \xv \approx 0.04\,\mg/\mz,
\ee
where $\mg$ is the total cold gas mass and $\mz$ is the mass of metals in cold gas. Eq.~(\ref{eqxsim}) only relates cold gas properties to each other and therefore is more fundamental than Eq.~(\ref{eqxoh}).

To evaluate Eq.~(\ref{eqxsim}) for each galaxy in the simulation, we used the cold gas metal masses $\mz$ given in the DeLucia-catalog. Those masses are reasonably accurate as demonstrated by \citet{DeLucia2004} and \citet{Croton2006} through a comparison of the simulated stellar mass--metallicity relation to the empirical mass--metallicity relation obtained from 53,000 star forming galaxies in the Sloan Digital Sky Survey \citep{Tremonti2004}. For most galaxies at $z=0$ the simulation yields metal fractions $\mz/\mg\approx0.01-0.04$ in the local Universe, thus implying $\xv\approx1-4$ in agreement with observed values \citep[e.g.][]{Boselli2002}.

\begin{figure}[h]
  \includegraphics[width=\columnwidth]{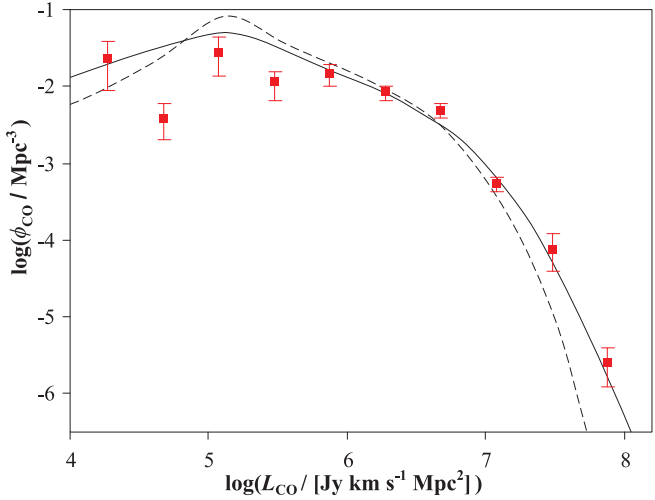}
  \caption{Luminosity function of CO(1--0)-emission (CO-LF) in the local Universe. The solid line represents the simulated CO-LF, obtained using the variable conversion factor $\xv$ of Eq.~(\ref{eqxsim}), and the dashed line represents the CO-LF, obtained using the constant conversion factor $\xc$ of Eq.~(\ref{eqxconst}). Square dots and error bars represent the empirical CO-LF determined by \citet{Keres2003}.}
  \label{fig_local_colf}
\end{figure}

Fig.~\ref{fig_local_colf} displays the simulated CO-LF for the variable $X$-factor $\xv$ (solid line) and the constant $X$-factor $\xc$ (dashed line) together with the empirical CO-LF \citep{Keres2003}, adjusted to $\h=0.73$. The comparison supports the variable $X$-factor of Eq.~(\ref{eqxsim}) against $\xc=2$ (and the same conclusion is found for other constant values of $\xc$). Using Eq.~(\ref{eqxsim}) also has the advantage that the cosmic evolution of the $X$-factor due to the evolution of metallicity is implicitly accounted for. Nevertheless Eq.~(\ref{eqxsim}) may not be appropriate at high redshift as discussed in Section \ref{section_discussion_evolution}.

\section{Cold gas disk sizes}\label{section_sizes}

Using the axially symmetric surface density profiles for \ha~and \hm~given in Eqs.~(\ref{eqsigmaha}, \ref{eqsigmahm}), we can define the \ha-radius $\rha$ and \hm-radius $\rhm$ of an axially symmetric galaxy as the radii corresponding to a detection limit $\Sigmanull$, i.e.
\bea
  \Sigmaha(\rha) & \equiv & \Sigmanull\,, \label{defrha} \\
  \Sigmahm(\rhm) & \equiv & \Sigmanull\,. \label{defrhm}
\eea
In this paper, we chose $\Sigmanull=1\,\msun{\rm pc}^{-2}$, corresponding to the deep survey of the Ursa Major group by \citet[e.g.][]{Verheijen2001}, but any other value could be adopted. In general Eqs.~(\ref{defrha}, \ref{defrhm}) do not have explicit closed-form solutions and must be solved numerically for each galaxy.

\begin{figure}[h]
  \includegraphics[width=\columnwidth]{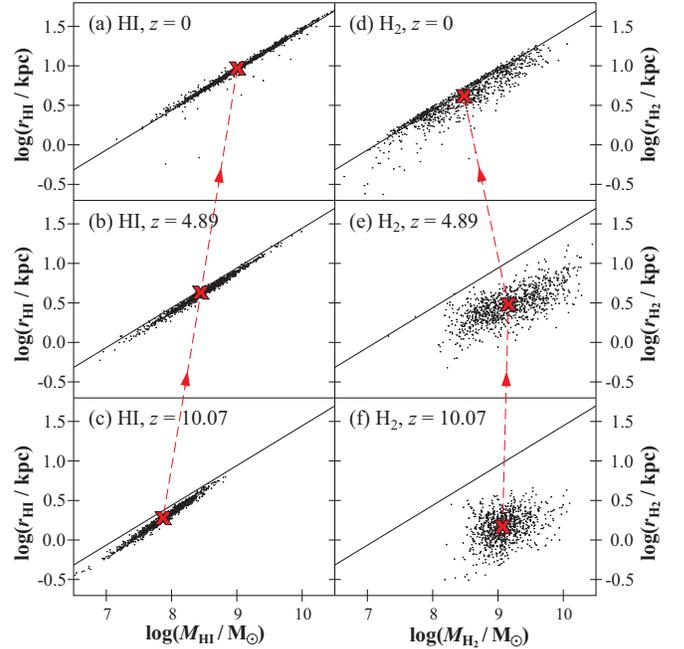}
  \caption{Simulated mass--radius relations for \ha~and \hm~at redshifts $z=0, 4.89, 10.07$, corresponding to the simulation snapshots 63, 21, 12. Black dots represent $10^3$ simulated galaxies and the solid lines show the power-law regression for the data in Fig.~\ref{fig_radius_evolution}a (i.e.~\ha~at $z=0$). The red crosses represent the cold gas mass-weighted averages of ($\mha$,$\rha$) and ($\mhm$,$\rhm$) in the simulation at each of the three redshifts.}
  \label{fig_radius_evolution}
\end{figure}

\begin{figure}[h]
  \includegraphics[width=\columnwidth]{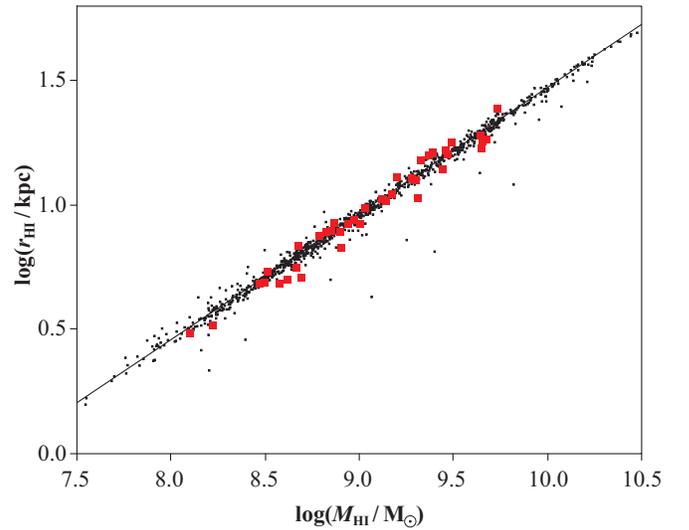}
  \caption{Relation between \ha-mass $\mha$ and \ha-radius $\rha$ for galaxies at redshift $z=0$. The black dots represent $10^3$ simulated galaxies and the solid line their linear regression. The slope of this power-law is 0.5, thus indicating a universal average \ha-surface density for all disk galaxies. The simulated data are identical to those plotted in Fig.~\ref{fig_radius_evolution}a. Red squares show measurements in the Ursa Major group by \citet{Verheijen2001}, who used the same definition of $\rha$ as this paper.}
  \label{fig_mhi_rhi}
\end{figure}

Results for $\rha$ and $\rhm$ at three epochs are displayed in Fig.~\ref{fig_radius_evolution}. Each graph shows $10^3$ simulated galaxies, drawn randomly from the catalog with a probability proportional to their cold gas mass. This selection rule ensures that rare objects at the high end of the MF are included. The arithmetic average of the points in each graph can be interpreted as the cold gas mass-weighted average of the displayed quantities. This average is marked in each graph to emphasize changes with redshift. The data in Fig.~\ref{fig_radius_evolution}a are shown again in Fig.~\ref{fig_mhi_rhi} together with measurements of 39 spiral galaxies in the Ursa Major group \citep[][]{Verheijen2001}.

Figs.~\ref{fig_radius_evolution} and \ref{fig_mhi_rhi} reveal several features, which we shall discuss hereafter: (i) the mass--radius relation for \ha~is a nearly perfect power-law with surprisingly small scatter; (ii) in general, radii become smaller with increasing redshift; (iii) the evolution of the mass--radius relation is completely different for \ha~and \hm.

The first result, i.e.~the strict power-law relation between $\mha$ and $\rha$, is strongly supported by measurements in the Ursa Major group \citep[][see Fig.~\ref{fig_mhi_rhi}]{Verheijen2001}. The best power-law fit to the simulation is
\be\label{eqmharhapowerlaw}
  \frac{\mha}{\msun} = 12\cdot\left(\frac{\rha}{\rm pc}\right)^{2.0}.
\ee
The rms-scatter of the simulated data around Eq.~(\ref{eqmharhapowerlaw}) is $\sigma=0.03$ in log-space, while the rms-scatter of the observations by \citet{Verheijen2001} is $\sigma=0.06$. This small scatter is particularly surprising as the more fundamental relation between $\mbary$ and $\rdstars$ shown in Fig.~\ref{fig_mass_scaleradius} exhibits a much larger scatter of $\sigma=0.26$. The square-law form of the power-law in Eq.~(\ref{eqmharhapowerlaw}) implies that the average \ha-surface density inside the radius $\rha$ is nearly identical for all galaxies, which have most of their \ha-mass inside the radius $\rha$,
\be
  \langle\Sigmaha\rangle_{r\leq\rha}\approx\frac{\mha}{\rha^2\pi}\approx3.8\,\msun~\rm pc^{-2}.
\ee
The existence of such a constant average density of \ha~can to first order be interpreted as a consequence of the fact that \ha~transforms into \hm~and stars as soon as its density and pressure are raised. In fact, observations show that $\Sigmaha$ saturates at about $6-10\,\msun~\rm pc^{-2}$ \citep{Blitz2006,Leroy2008} and that higher cold gas densities are generally dominated by $\Sigmahm$. Therefore, \ha~maintains a constant surface density during the evolution of any isolated galaxy as long as enough \ha~is supplied from an external source, e.g.~by cooling from a hot medium as assumed in the recipes of the DeLucia-catalog. This also explains why the power-law relation between $\mha$ and $\rha$ remains nearly constant towards higher redshift in the simulation (Fig.~\ref{fig_radius_evolution}a--c). About 1\% of the simulated galaxies at redshift $z=0$ lie far off the power-law relation (i.e.~are outside 5-$\sigma$ of the best fit), typically towards smaller radii (see Fig.~\ref{fig_mhi_rhi}). One might first expect that these objects have a higher \ha-surface density, while, in fact, the contrary applies. These galaxies have very flat \ha-profiles with most of the \ha-mass lying outside the radius $\rha$, and therefore they would require a lower sensitivity limit than $1\,\msun~\rm pc^{-2}$ for a useful definition of $\rha$. Such galaxies are indeed very difficult to map due to observational surface brightness limitations.

The radii $\rha$ and $\rhm$ become smaller towards higher redshift. This is a direct consequence of the cosmic evolution of the virial radii $\rvir$ of the haloes in the Millennium Simulation, which affects the disk scale radius $\rdstars$ in Eq.~(\ref{eqrd}). As shown by \citet{Mo1998}, $\rvir$ scales as $(1+z)^{-1.5}$ for a fixed circular velocity or as $(1+z)^{-1}$ for a fixed halo mass, consistent with high-redshift observations ($z=2.5-6$) in the Hubble Ultra Deep Field (UDF) by \citet{Bouwens2004}. Their selection criteria include all but the reddest starburst galaxies in the UDF and some evolved galaxies. It should be emphasized that the phenomenological size evolution of galaxies is not properly understood, and even modern $N$-body/SPH-simulation cannot yet accurately reproduce the sizes of galaxies.

In analogy to the mass--radius power-law relation for \ha, our simulation predicts a similar relation, again nearly a square-law, for \hm~at redshift $z=0$ (see Fig.~\ref{fig_radius_evolution}d). This power-law is consistent with observations of the two face-on spiral galaxies M\,51 \citep{Schuster2007} and NGC\,6946 \citep{Crosthwaite2007}. For the small \hm/\ha-ratios found in the local Universe the $\mhm$--\,$\rhm$ relation is linked to the $\mha$--\,$\rha$ relation, because both $\mhm$ and $\rhm$ can be regarded as a fraction $(<1)$ of, respectively, $\mha$ and $\rha$. However, there is no fundamental reason for a constant surface density of \hm~and the smaller sizes of high-redshift galaxies implies a higher pressure of the ISM and thus a much higher molecular fraction by virtue of Eqs.~(\ref{eqfc}, \ref{eqfg}). Therefore, \hm-masses become uncorrelated to \ha~and tend to increase with redshift out to $z\approx5$, while $\rhm$ decreases. Hence, the $\mhm$--\,$\rhm$ relation must move away from the power-law $\mhm\sim\rhm^2$ found at $z=0$ (see Figs.~\ref{fig_radius_evolution}d--f).

\section{Realistic velocity profiles}\label{section_velocities}

In this section we derive circular velocity profiles and atomic and molecular-radio line profiles for the simulated galaxies in the DeLucia-catalog. Circular velocity profiles $\vc(r)$ for various galaxies are derived over the Sections \ref{section_vchaloe}--\ref{section_vcbulge} and transcribed to radio line profiles for edge-on galaxies in Section \ref{section_radio_lines}. Results for the local and high-redshift Universe are presented in Section \ref{section_line_evolution}.

\subsection{Velocity profile of a spherical halo}\label{section_vchaloe}

To account for the narrowness of the emission lines observed in the central gas regions of many galaxies \citep[e.g.][]{Sauty2003}, we require a halo model with vanishing velocity at the center, as opposed to, for example, the commonly adopted singular isothermal sphere with a density $\rhoh(r)\sim r^{-2}$ and a constant velocity profile. We chose the Navarro--Frenk--White \citep[NFW,][]{Navarro1995,Navarro1996} model, which relies on high resolution numerical simulations of dark matter haloes in equilibrium. These simulations revealed that haloes of all masses in a variety of dissipation-less hierarchical clustering models are well described by the spherical density profile
\be\label{rhonfw}
  \rhoh(r) = \rho_0\,{\Big[(r/\rh)(1+r/\rh)^2\Big]}^{-1},
\ee
where $\rho_0$ is a normalization factor and $\rh$ is the characteristic scale radius of the halo. This profile is also supported by the Hubble Space Telescope analysis of the weak lensing induced by the galaxy cluster MS 2053-04 at redshift $z=0.58$ \citep{Hoekstra2002}. $\rhoh(r)$ varies as $r^{-1}$ at the halo center and continuously steepens to $r^{-3}$ for $r\rightarrow\infty$. It passes through the equilibrium profile of the self-gravitating isothermal sphere, i.e.~$\rhoh(r)\sim r^{-2}$, at $r=\rh$.

The definition of $\rvir$ in the Millennium Simulation given in Eq.~(\ref{eqmvir}) implies that
\be\label{eqrhonull}
  \rho_0 = \frac{200}{3}\,\frac{\rhoc\,\ch^3}{\ln(1+\ch)-\ch/(1+\ch)},
\ee
where $\ch\equiv\rvir/\rh$ is referred to as the halo concentration parameter. Most numerical models predict that $\ch$ scales with the virial mass $\mvir$, defined as the mass inside the radius $\rvir$, according to a power-law \citep[e.g.][]{Navarro1997,Bullock2001,Dolag2004,Hennawi2007}. Here we shall use the result of \citet{Hennawi2007},
\be\label{eqch}
  \ch =\frac{12.3}{1+z}\left(\frac{\mvir}{1.3\cdot10^{13}\,\h^{-1}\msun}\right)^{-0.13},
\ee
which is consistent with recent empirical values of the matter concentration in galaxy clusters derived from X-ray measurements and strong lensing data \citep{Comerford2007}.

For a spherical halo, the circular velocity profile is given by ${\vch}^2(r)=G\mh(r)/r$ with $\mh(r)=4\pi\int_0^r{\rm d}\tilde{r}\,\tilde{r}^2\,\rhoh(\tilde{r})$. Using Eqs.~(\ref{rhonfw}, \ref{eqrhonull}), this implies that
\be\label{eq_vch}
  {\vch}^2(x) = \frac{G\mvir}{\rvir}\times\frac{\ln(1+\ch\x)-\frac{\ch\x}{1+\ch\x}}{\x[\ln(1+\ch)-\frac{\ch}{1+\ch}]},
\ee
where $\x\equiv r/\rvir$ (thus $\ch\x=r/\rh$). This velocity vanishes at the halo center, then climbs to a maximal value $V_{\rm max}=1.65\,\rh\sqrt{G\rho_0}$ at $r=2.16\,\rh$, from where it decreases monotonically with $r$, typically reaching $0.65-0.95\,V_{\rm max}$ at $r=\rvir$ ($\x=1$) with the extremes corresponding, respectively, to $\ch=25$ and $\ch=5$. For larger radii, the velocity asymptotically approaches the point-mass velocity profile ${\vch}^2(r)=G\mvir/r$.

\subsection{Velocity profile of a flat disk}\label{section_vcdisk}
For simplicity, we assume in this section that the galactic disk is described by a single exponential surface density for stars and cold gas,
\be\label{eqsigmad}
  \Sigmad(r) = \frac{\md}{2\pi\,\rd^2}\,\exp\left(-\frac{r}{\rd}\right),
\ee
where $\md$ is the total disk mass, taken as the sum of the cold gas mass and the stellar mass in the disk. In most real galaxies the stellar surface densities, are slightly more compact (see Section \ref{section_subdivision}), but we found that including this effect does not significantly modify the shape of the atomic and molecular emission lines. In fact, the radius, which maximally contributes to the disk mass, i.e.~the maximum of $r\,\Sigmad(r)$, is $r=\rd$. Therefore, we expect the gravitational potential to differ significantly from the point-mass potential only for $r$ of order $\rd$ or smaller. Applying Poisson's equation to the surface density of Eq.~(\ref{eqsigmad}), the gravitational potential in the plane of the disk becomes
\be\label{diskpotential}
  \potd(r) = -\frac{G\md}{2\pi\rd^2}\int\!\!\!\int_D\frac{\exp(-\tilde{r}/\rd)\,\tilde{r}\,{\rm d}\tilde{r}\,{\rm d}\theta}{(r^2{+\tilde{r}^2}{-2}r\tilde{r}\cos\theta)^{1/2}},
\ee
where the integration surface $D$ is given by $\tilde{r}\in[0,\infty)$, $\theta=[0,2\pi)$.

The velocity profile for circular orbits in the plane of the disk can be calculated as ${\vcd}^2=r\,{\rm d}\potd/{\rm d}r$. The integral in Eq.~(\ref{diskpotential}) is elliptic, and hence there are no exact closed-from expressions for $\potd$ and $\vcd$. However, in this study we numerically found that an excellent approximation is given by
\be\label{eq_vcd}
  {\vcd}^2(x) \approx \frac{G\md}{\rvir}\times \qquad\qquad\qquad\qquad\qquad\quad
\ee
\vspace{-0.3cm}
\be
  \frac{\cd+4.8\cd\exp[-0.35\cd\x-3.5/(\cd\x)]}{\cd\x+(\cd\x)^{-2}+2(\cd\x)^{-1/2}},\quad \nonumber
\ee
where $\cd\equiv\rvir/\rd$ is the disk concentration parameter in analogy to the halo concentration parameter $\ch$ of Section \ref{section_vchaloe}. Eq.~(\ref{eq_vcd}) is accurate to less than 1\% over the whole range $r=0-10\,\rd$ and it correctly converges towards the circular velocity of a point-mass potential, ${\vch}^2(r)=G\md/r$, for $r\rightarrow\infty$.

Like in Section \ref{section_appl}, we shall use Eq.~(\ref{eqrd}) to compute $\rd$ and $\cd$ in the DeLucia-catalog. This approach is slightly inconsistent because Eq.~(\ref{eqrd}) was derived by \citet{Mo1998} under the assumption that the disk is supported by an isothermal halo with $\rhoh(r)\sim r^{-2}$, while in Section \ref{section_vchaloe} we have assumed the more complex NFW-profile. We argue, however, that Eq.~(\ref{eqrd}) with the empirical correction of Eq.~(\ref{eqxi}) is sufficiently accurate, as it successfully reproduces the observed relation between $\mbary$ and $\rd$ (see Fig.~\ref{fig_mass_scaleradius}) as well as the relation between $\mha$ and $\rha$ (see Fig.~\ref{fig_mhi_rhi}). It can also be shown that, for realistic values of the halo concentration $\ch$ ($10-20$ for one-galaxy systems) and the spin parameter $\lambda$ ($0.05-0.1$, e.g.~\citealp{Mo1998}), the scale radius of the halo $\rh$ and the disk radius $\rd$ are similar. Hence, the main mass contribution of the disk comes indeed from galactocentric radii, where the halo profile is approximately isothermal, thus justifying the assumption made by \cite{Mo1998} to derive Eq.~(\ref{eqrd}).

\subsection{Velocity profile of the bulge}\label{section_vcbulge}

Many models for the surface brightness or surface density profiles of bulges have been proposed \citep[e.g.~overview by][]{Balcells2001}. A rough consensus seems established that no single surface density profile can describe a majority of the observed bulges, but that they are generally well matched by the class of S\'{e}rsic-functions \citep{Sersic1968}, $\Sigmab(r)\sim\exp[-(r/\rb)^{1/n}]$, where the exponent $n$ depends on the morphological type \citep{Andredakis1995}, such that $n\approx4$ for lenticular/early-type galaxies (de Vaucouleurs-profile) and $n\approx1$ for the bulges of late-type galaxies (exponential profile). \cite{Courteau1996} find slightly steeper profiles with $n=1-2$ for nearly all spirals in a sample of 326 spiral galaxies using deep optical and IR photometry, and they show that by imposing $n=1$ for all late-type galaxies, the ratio between the exponential scale radius of the bulge $\rb$ and the scale radius of the disk $\rd$ is roughly constant, $\rb\approx0.1\,\rdstars\approx0.05\,\rdgas$. We shall therefore assume that all bulges have an exponential projected surface density,
\be
  \Sigmab(r)=\frac{\mb}{2\pi\,\rb^2}\,\exp\left(-\frac{r}{\rb}\right)
\ee
with $\rb=0.05\,\rd$.

For simplicity, we assume that the bulges of all galaxies are spherical and thus described by a radial space density function $\rhob(r)$. This function is linked to the projected surface density via $\Sigmab(r)=\int_{-\infty}^{\infty}{\rm d}z\,\rhob\big[(r^2+z^2)^{1/2}]$. Numerically, we find that this model for $\rhob(r)$ is closely approximated by the Plummer model \citep{Plummer1911}, more often used in the context of clusters,
\be\label{eqrhob}
  \rhob(r) \approx \frac{3\,\mb}{4\pi\,\rbp^3}\left[1+\left(\frac{r}{\rbp}\right)^2\right]^{-5/2},
\ee
with a characteristic Plummer radius $\rbp\approx1.7\,\rb$. The circular velocity profile $\vcb$ corresponding to Eq.~(\ref{eqrhob}) is given by ${\vcb}^2(r)=G\mb(r)/r$ with $\mb(r)=4\pi\int_0^r{\rm d}\tilde{r}\,\tilde{r}^2\,\rhob(\tilde{r})$. This solves to
\be\label{eq_vcb}
  {\vcb}^2(x) = \frac{G\mb}{\rvir}\times\frac{(\cb x)^2\cb}{\left[1+(\cb x)^2\right]^{3/2}},
\ee
where $\cb\equiv\rvir/\rbp\approx 12\,\cd$ is the bulge concentration parameter.

\subsection{Line shapes from circular velocities}\label{section_radio_lines}

\begin{figure}[h]
  \includegraphics[width=\columnwidth]{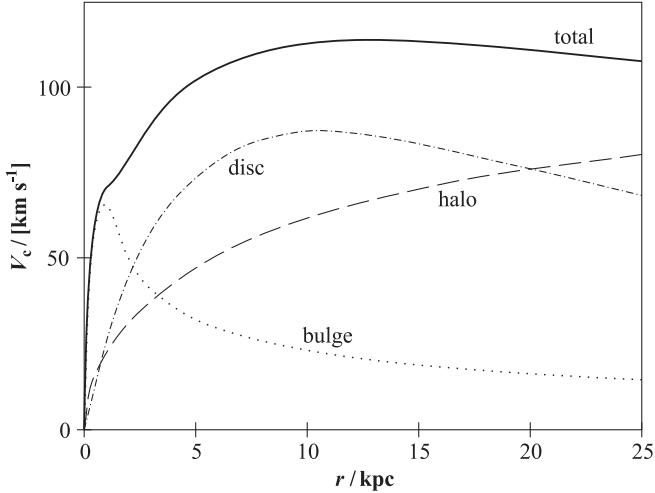}
  \caption{Circular velocity profile of a typical simulated galaxy with a small bulge at redshift $z=0$. The total circular velocity (solid line) is given by the circular velocity of the halo (dashed line), the disk (dash-dotted line), and the bulge (dotted line) via Eq.~(\ref{eqvctot}).}
  \label{fig_velocity_example}
\end{figure}

The addition rule for gravitational potentials implies that the circular velocity profile of the combined halo--disk--bulge system in the plane of the disk is given by
\be\label{eqvctot}
  \vc^2(x)={\vch}^2(x)+{\vcd}^2(x)+{\vcb}^2(x),
\ee
where $x\equiv r/\rvir$ as in Sections \ref{section_vchaloe}--\ref{section_vcbulge}. According to Eqs.~(\ref{eq_vch}, \ref{eq_vcd}, \ref{eq_vcb}), this profile is determined by six parameters: the three form-parameters $\ch$, $\cd$, $\cb$ and the three mass-scales $\mvir/\rvir$, $\md/\rvir$, $\mb/\rvir$. The form-parameters were calculated as explained in Sections \ref{section_vchaloe}--\ref{section_vcbulge}, while the mass-scales were directly adopted from the DeLucia-catalog. For the satellite galaxies with no resolved halo (see Section \ref{section_background}), $\mvir$ and $\rvir$ were approximated as the corresponding quantities of the original galaxy halo just before its disappearance. An exemplar circular velocity profile for a galaxy in the DeLucia-catalog at redshift $z=0$ is shown in Fig.~\ref{fig_velocity_example}.

\begin{figure}[h]
  \begin{center}
    \includegraphics[width=6cm]{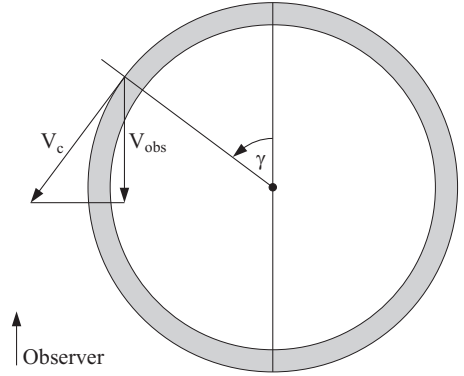}
    \caption{Apparent velocity $\vo$ induced by the infinitesimal ring element at the angle $\gamma$.}
    \label{fig_ring_model}
  \end{center}
\end{figure}

\begin{figure}[h]
  \includegraphics[width=\columnwidth]{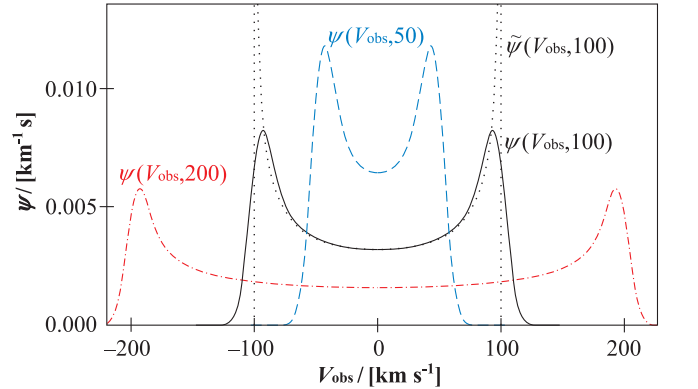}
  \caption{Illustration of the functions $\psi$ (Eq.~\ref{eqpsi}) and $\tilde{\psi}$ (Eq.~\ref{eqpsitilde}), which represent the normalized emission line of a homogeneous edge-on disk or ring with constant circular velocity.}
  \label{fig_psi_function}
\end{figure}

In order to evaluate the profile of a radio emission line associated with any velocity profile $\vc(r)$, we shall first consider the line profile of a homogeneous flat ring with constant circular velocity $\vc$ and a total luminosity of unity. If a point of the ring is labeled by the angle $\gamma$ it forms with the line-of-sight (see Fig.~\ref{fig_ring_model}), the apparent projected velocity of that point is given by $\vo=\vc\,\sin\gamma$. The ensemble of all angles $\gamma\in[0,2\pi)$ therefore spans a continuum of apparent velocities $\vo\in(-\vc,\vc)$ with a luminosity density distribution $\tilde{\psi}(\vo,\vc)\sim{\rm d}\gamma/{\rm d}\vo$. Imposing the normalization condition $\int{\rm d}\vo\tilde{\psi}(\vo)=1$, we find that the edge-on line profile of the ring is given by
\be\label{eqpsitilde}
  \tilde{\psi}(\vo,\vc) = \left\{\begin{array}{ll}
  \frac{1}{\pi\sqrt{\vc^2-\vo^2}} & {\rm if~}|\vo|<\vc \\
  0 & \rm otherwise.
  \end{array}\right.
\ee
This profile exhibits spurious divergent singularities at $|\vo|\rightarrow\vc$, which, in reality, are smoothed by the random, e.g.~turbulent, motion of the gas. We assume that this velocity dispersion is given by the constant $\vg=8\rm\,km~s^{-1}$, which is consistent with the velocity dispersions observed across the disks of several nearby galaxies \citep[e.g.][]{Shostak1984,Dickey1990,Burton1971}. The smoothed velocity profile is then given by
\be\label{eqpsi}
  \psi(\vo,\vc)\!=\!\frac{\vg^{-1}}{\!\sqrt{2\pi}}\!\int\!\!{\rm d}V\!\exp\!\!\bigg[\!\frac{(\vo\!\!-\!V)^2}{-2\vg^2}\bigg]\tilde{\psi}(V,\vc),
\ee
which conserves the normalization $\int{\rm d}\vo\psi(\vo)=1$. Some examples of the functions $\tilde{\psi}(\vo,\vc)$ and $\psi(\vo,\vc)$ are plotted in Fig.~\ref{fig_psi_function}.

From the edge-on line profile $\psi(\vo,\vc)$ of a single ring and the face-on surface densities of atomic and molecular gas, $\Sigmaha(r)$ and $\Sigmahm(r)$, we can now evaluate the edge-on profiles of emission lines associated with the entire \ha- and \hm-disks, respectively. Since \hm-densities are most commonly inferred from CO-detections, we shall hereafter refer to all molecular emission lines as ``the CO-line''. The edge-on line profiles (or ``normalized luminosity densities'') $\Psi_{\rm\ha}(\vo)$ and $\Psi_{\rm CO}(\vo)$ are given by
\bea\label{eqfullline}
  \Psi_{\rm\ha}(\vo) & = & \frac{2\pi}{\mha}\int_0^\infty\!\!\!\!{\rm d}r\,r\,\Sigmaha(r)\,\psi\big(\vo,\vc(r)\big), \\
  \Psi_{\rm CO}(\vo) & = & \frac{2\pi}{\mhm}\int_0^\infty\!\!\!\!{\rm d}r\,r\,\Sigmahm(r)\,\psi\big(\vo,\vc(r)\big).
\eea
These two functions satisfy the normalization conditions $\int{\rm d}\vo\Psi_{\rm\ha}(\vo)=1$ and $\int{\rm d}\vo\Psi_{\rm CO}(\vo)=1$. To obtain intrinsic luminosity densities, $\Psi_{\rm\ha}(\vo)$ must be multiplied by the integrated luminosity of the \ha-line [see Eq.~(\ref{eq_mha})] and $\Psi_{\rm CO}(\vo)$ must be multiplied by the integrated luminosity of the considered molecular emission line, e.g.~the integrated luminosity of the CO(1--0)-line given in Eq.~(\ref{eq_mhm}).

\begin{figure}[h]
  \includegraphics[width=\columnwidth]{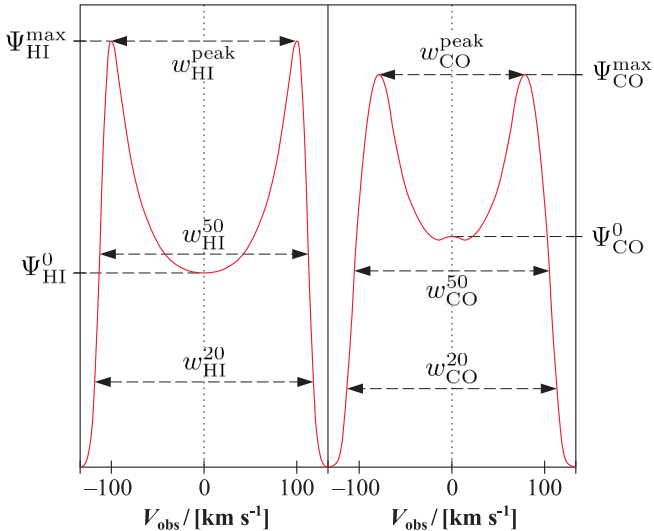}
  \caption{Simulated edge-on \ha- and CO-emission lines for the exemplar galaxy, for which the circular velocity profile is shown in Fig.~\ref{fig_velocity_example}. The line profiles have been computed using Eq.~(\ref{eqfullline}).}
  \label{fig_line_example}
\end{figure}

Fig.~\ref{fig_line_example} displays the line profiles $\Psi_{\rm \ha}(\vo)$ and $\Psi_{\rm CO}(\vo)$ for the exemplar galaxy with the velocity profile shown in Fig.~\ref{fig_velocity_example}. All line profiles produced by our model are mirror-symmetric, but they can, in principle, differ significantly from the basic double-horned function $\psi(\vo)$. Especially for CO, where the emission from the bulge can play an important role, several local maxima can sometimes be found in the line profile, in qualitative agreement with various observations \citep[e.g.][]{Lavezzi1998}.

\subsection{Results and discussion}\label{section_line_evolution}

For every galaxy in the DeLucia-catalog, we computed the edge-on line profiles $\Psi_{\rm \ha}(\vo)$ and $\Psi_{\rm CO}(\vo)$, from which we extracted the line parameters indicated in Fig.~\ref{fig_line_example}. $\lhacenter\equiv\Psi_{\rm \ha}(0)$ and $\lcocenter\equiv\Psi_{\rm CO}(0)$ are the luminosity densities at the line center, and $\lhamax$ and $\lcomax$ are the peak luminosity densities, i.e.~the absolute maxima of $\Psi_{\rm \ha}(\vo)$ and $\Psi_{\rm CO}(\vo)$. $\whapeak$ and $\wcopeak$ are the line widths measured between the left and the right maximum. These values vanish if the line maxima are at the line center, such as found, for example, in slowly rotating systems. $\whafifty$, $\wcofifty$, $\whatwenty$, and $\wcotwenty$ are the line widths measured at, respectively, the 50-percentile level or the 20-percentile level of the peak luminosity densities -- the two most common definitions of observed line widths.

We shall now check the simulated line widths against observations by analyzing their relation to the mass of the galaxies. Here, we shall refer to all line width versus mass relations as Tully--Fisher relations (TFRs), since they are generalized versions of the original relation between line widths and optical magnitudes of spiral galaxies \citep{Tully1977}. A variety of empirical TFRs have been published, such as the stellar mass-TFR and the baryonic-TFR \citep{McGaugh2000}. The latter relates the baryon mass (stars$+$gas) of spiral disks to their line widths (or circular velocities) and is probably the most fundamental TFR detected so far, obeying a single power-law over five orders of magnitude in mass. We have also investigated the less fundamental empirical TFR between $\mha$ and $\whafifty$ -- hereafter the HI-TFR -- using the spiral galaxies of the HIPASS catalog. Assuming no prior knowledge on the inclinations of the HIPASS-galaxies, but taking an isotropic distribution of their axes as given, we found that the most-likely relation is
\be\label{eqhitfr}
  \log\left(\frac{\mha}{\msun}\right) = 2.86+2.808\cdot\log\left(\frac{\whafifty}{\rm km~s^{-1}}\right)
\ee
for the Hubble parameter $\h=0.73$. Relative to Eq.~(\ref{eqhitfr}) the HIPASS data exhibit a Gaussian scatter with $\sigma=0.38$ in $\log(\mha)$. Our method to find Eq.~(\ref{eqhitfr}) will be detailed in a forthcoming paper (Obreschkow et al.~in prep.), especially dedicated to the HI-TFR.

\begin{figure*}
  \includegraphics[width=17.5cm]{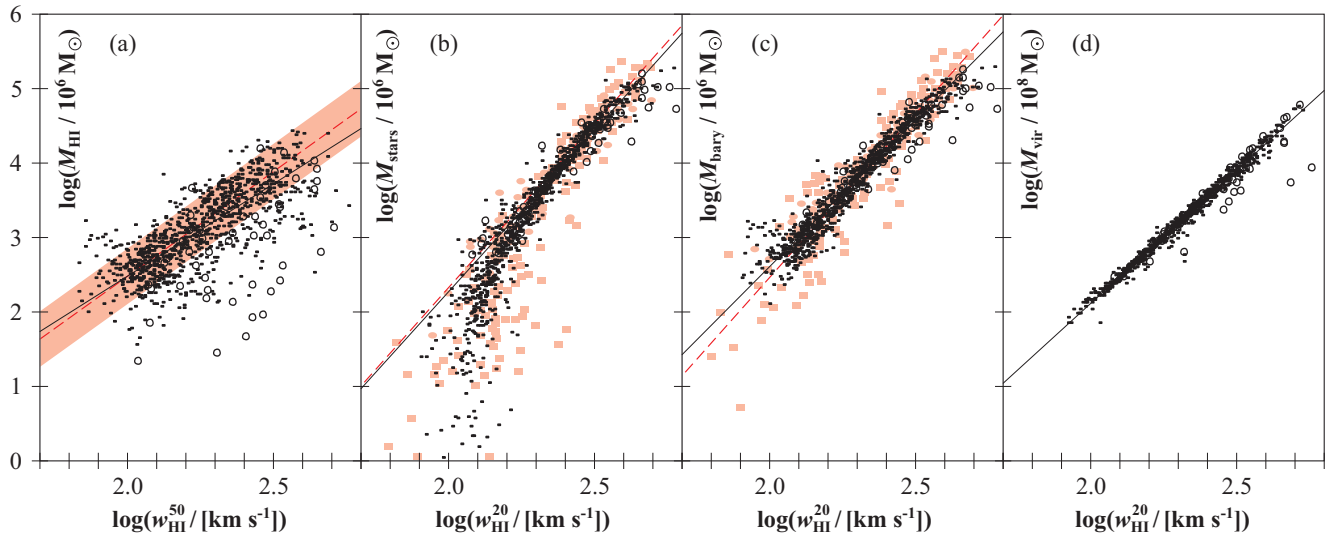}
  \caption{Relations between edge-on line widths and different mass tracers for the local Universe. $10^3$ simulated galaxies are represented by black dots (spiral galaxies) and black circles (elliptical galaxies). The solid lines represent power-law fits to the simulated spiral galaxies; their slopes are respectively $\alpha_{\rm HI}=2.5$, $\alpha_{\rm stars}=4.3$, $\alpha_{\rm bary}=3.9$, and $\alpha_{\rm vir}=3.6$. In case of Fig.~\ref{fig_alltfr_local}b, this fit only includes galaxies with $\ms>10^9\,\msun$. Fig.~\ref{fig_alltfr_local}d does not include satellite galaxies without haloes (see Section \ref{section_background}), for which $\mvir$ is poorly defined. The dashed red line and shaded zone in Fig.~\ref{fig_alltfr_local}a represent our observational determination and 1-$\sigma$ scatter of the \ha-TFR from the HIPASS data (see Section \ref{section_line_evolution}). The rose dots and dashed lines in Figs.~\ref{fig_alltfr_local}b, c are the observational data and power-law regressions from \citet{McGaugh2000} and references therein; this sample include low surface brightness galaxies. The slopes of these empirical power-laws are respectively $\alpha_{\rm HI}=2.8$ (see Eq.~\ref{eqhitfr}), $\alpha_{\rm stars}=4.4$, and $\alpha_{\rm bary}=4.4$.}
  \label{fig_alltfr_local}
\end{figure*}

Figs.~\ref{fig_alltfr_local}a--d show four TFRs at redshift $z=0$. Each figure represents $10^3$ simulated galaxies (black), randomly drawn from the simulation with a probability proportional to their cold gas mass in order to include the rare objects in the high end of the MF. Spiral and elliptical galaxies are distinguished as black dots and open circles.

Fig.~\ref{fig_alltfr_local}a shows the simulated HI-TFR together with the empirical counterpart given in Eq.~(\ref{eqhitfr}). This comparison reveals good consistency between observation and simulation for spiral galaxies. However, the elliptical galaxies lie far off the HI-TFR. In fact, simulated elliptical galaxies generally have a significant fraction of their cold hydrogen in the molecular phase, consistent with the galaxy-type dependence of the \hm/\ha-ratio first identified by \cite{Young1989b}. Therefore, \ha~is a poor mass tracer for elliptical galaxies, both in simulations and observations, leading to their offset from the TFR when only \ha-masses are considered. There seems to be no direct analog to the HI-TFR for elliptical galaxies.

Figs.~\ref{fig_alltfr_local}b, c respectively display the simulated stellar mass-TFR and the baryonic-TFR, together with the observed data of \citet{McGaugh2000} corrected for $\h=0.73$. These data include various galaxies from dwarfs to giant spirals, whose edge-on line widths were estimated from the observed ones using the inclinations determined from the optical axis ratios. Figs.~\ref{fig_alltfr_local}b, c reveal a surprising consistency between simulation and observation. In Fig.~\ref{fig_alltfr_local}b, both the simulated and observed data show a systematic offset from the power-law relation for all galaxies with $\whatwenty\lesssim200\rm~km\,s^{-1}$. Yet, the power-law relation is restored as soon as the cold gas mass is added to the stellar mass (Fig.~\ref{fig_alltfr_local}c), thus confirming that the TFR is indeed fundamentally a relation between mass and circular velocity.

It is interesting to consider the prediction of the simulation for the most fundamental TFR, i.e.~the one between the total dynamical mass, taken as the virial mass $\mvir$, and the circular velocity, represented by the line width $\whatwenty$. This relation is shown in Fig.~\ref{fig_alltfr_local}d and reveals indeed a 2--3 times smaller scatter in $\log(\rm mass)$ than the baryonic-TFR, hence confirming its fundamental character.

Although the simulated elliptical galaxies shown in Figs.~\ref{fig_alltfr_local}b--d roughly align with the respective TFRs for spiral galaxies, their scatter is larger. This is caused by the mass-domination of the bulge, which leads to steep circular velocity profiles $\vc(r)$ with a poorly defined terminal velocity. Therefore, line widths obtained by averaging over the whole elliptical galaxy are weak tracers of its spin. This picture seems to correspond to observed elliptical galaxies, where the central line widths, corresponding to the velocity dispersion in the bulge dominated parts, are more correlated to the stellar mass than the line widths of the whole galaxy \citep[see Faber-Jackson relation, e.g.][]{Faber1976}.

\cite{Kassin2007} noted that $S_{0.5}\equiv(0.5\,\vc^2+\vg^2)^{1/2}$ is a better kinematic estimator than the circular velocity $\vc$, in the sense that it markedly reduces the scatter in the stellar mass-TFR. However, since our model assumes a constant gas velocity dispersion $\vg$ for all galaxies, it is not possible to investigate this estimator.

\begin{figure*}
  \includegraphics[width=17.5cm]{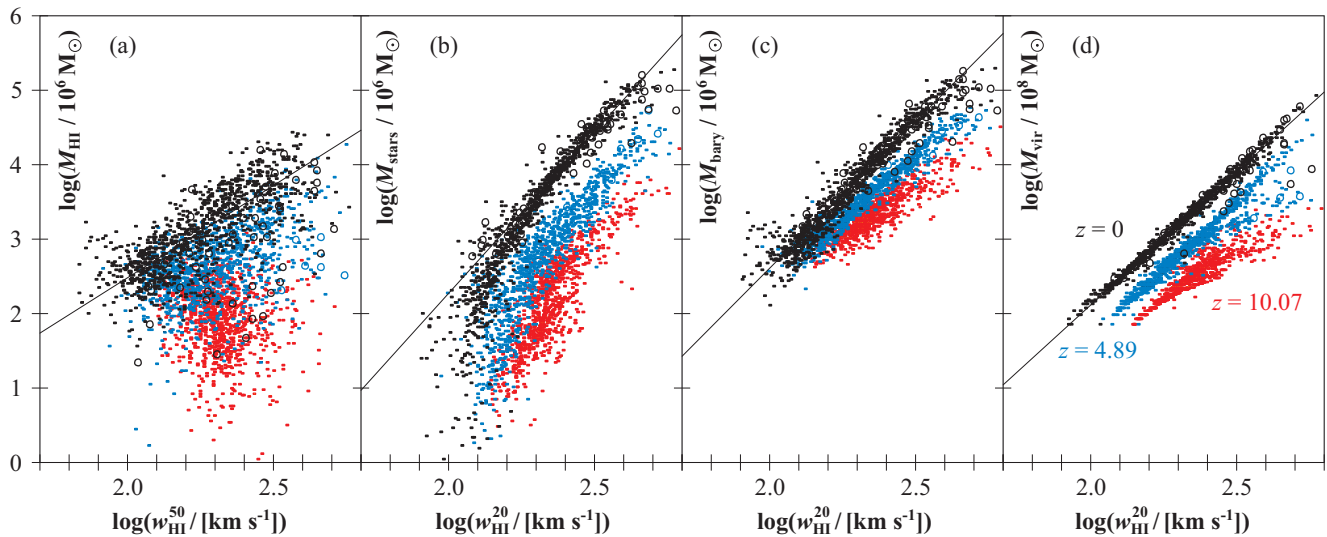}
  \caption{Simulated cosmic evolution of the different line width--mass relations shown in Fig.~\ref{fig_alltfr_local}. Spiral and elliptical galaxies are respectively represented by dots and circles. Black color corresponds to redshift $z=0$ (identically to Fig.~\ref{fig_alltfr_local}), while blue and red color respectively represent $z=4.89$ and $z=10.07$. The solid black lines are power-law fits to the spiral galaxies at $z=0$, where in case of Fig.~\ref{fig_alltfr_evolution}b only galaxies with $\ms>10^9\,\msun$ have been considered. The number of elliptical galaxies decreases with redshift -- a consequence of the merger- and instability-driven prescriptions for bulge formation in the DeLucia-catalog.}
  \label{fig_alltfr_evolution}
\end{figure*}

The predicted evolution of the four TFRs in Figs.~\ref{fig_alltfr_local}a--d is shown in Figs.~\ref{fig_alltfr_evolution}a--d. In all four cases, the simulation predicts two important features: (i) galaxies of identical mass (respectively $\mha$, $\ms$, $\mbary$, $\mvir$) have broader lines (and larger circular velocities) at higher redshift, and (ii) the scatter of the TFRs generally increases with redshift. The first feature is mainly a consequence of the mass--radius--velocity relation of the dark matter haloes assumed in the Millennium Simulation \citep[see][]{Croton2006,Mo1998}. This relation predicts that, given a constant halo mass, $\vc$ scales as $(1+z)^{1/2}$ for large $z$. Furthermore, the ratios $\mha/\mvir$ and $\ms/\mvir$ on average decrease with increasing redshift, explaining the stronger evolution found in Figs.~\ref{fig_alltfr_evolution}a, b relative to Figs.~\ref{fig_alltfr_evolution}c, d.

The increase of scatter in the TFRs with redshift is a consequence of the lower degree of virialization at higher redshifts, which, in the model, is accounted for via the spin parameter $\lambda$ of the haloes. $\lambda$ is more scattered at high redshift due to the young age of the haloes and the higher merger rates. More scatter in $\lambda$ leads to more scatter in the radius $\rd$ via Eq.~(\ref{eqrd}) and thus to more scatter in the circular velocity $\vc$ via Eqs.~(\ref{eq_vcd}, \ref{eq_vcb}).

\begin{figure}[h]
  \includegraphics[width=\columnwidth]{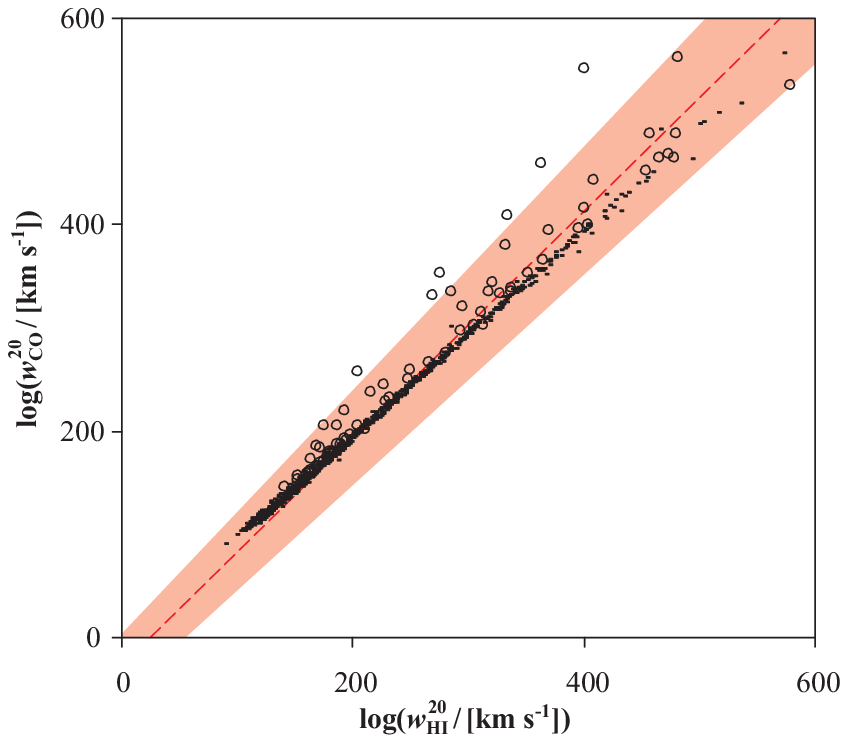}
  \caption{Relation between line widths of \ha~and CO. $10^3$ simulated galaxies are represented by black dots (spiral galaxies) and black circles (elliptical galaxies). The red dashed line and rose-shaded zone represent the best fit and its 1-$\sigma$ confidence interval to observations of 44 galaxies in clusters presented by \citet{Lavezzi1998}.}
  \label{fig_whi_wco}
\end{figure}

Current observational databases of resolved CO-line profiles are much smaller than \ha-databases and their signal/noise characteristics are inferior. Nevertheless efforts to check the use of CO-line widths for probing TFRs \citep[e.g.][]{Lavezzi1998} have led to the conclusion that in most spiral galaxies the CO-line widths are very similar to \ha-line widths, even though the actual line profiles may radically differ. Fig.~\ref{fig_whi_wco} shows our simulated relation between $\whatwenty$ and $\wcotwenty$, as well as the linear fit to observations of 44 galaxies in different clusters \citep{Lavezzi1998}. These observations are indeed consistent with the simulation. The simulated elliptical galaxies tend to have higher $\wcotwenty/\whatwenty$-ratios than the spiral ones, due to fast circular velocity of the bulge component implied by its own mass.

The line profiles and widths studied in this section correspond to galaxies observed edge-on. First order corrections for spiral galaxies seen at an inclination $i\neq90\rm\,deg$ can be obtained by dividing the normalized luminosity densities $\lhacenter$, $\lhamax$, $\lcocenter$, $\lcomax$ by $\sin i$, while multiplying the line widths  $\whafifty$, $\whatwenty$, $\whapeak$, $\wcofifty$, $\wcotwenty$, $\wcopeak$ by $\sin i$. More elaborate corrections, accounting for the isotropy of the velocity dispersion $\vg$, were given by \cite{Makarov1997} and \cite{Kannappan2002}.

\section{Discussion}\label{section_discussion}

We used a list of physical prescriptions to post-process the DeLucia-catalog and showed that many simulation results match the empirical findings from the local Universe. However, this approach raises two major questions: (i) Are the applied prescriptions consistent with the DeLucia-catalog in the sense that they represent a compatible extension of the semi-analytic recipes used by \cite{DeLucia2007} and \cite{Croton2006}? (ii) What are the limitations of our prescriptions at low and high redshifts?

\subsection{Consistency of the model}

The consistency question arises, because the DeLucia-catalog relies on a simplified version of a Schmidt--Kennicutt law \citep{Schmidt1959,Kennicutt1998}, i.e.~a prescription where the star formation rate (SFR) scales as some power of the surface density of the ISM. However, in a smaller-scaled picture, new stars are bred inside molecular clouds, and hence it must be verified whether our prescription to assign \hm-masses to galaxies is compatible with the macroscopic Schmidt--Kennicutt law. Our prescription exploited the empirical power-law between the pressure of the ISM and its molecular content, as first presented by \cite{Blitz2004,Blitz2006}. Based on this power-law relation, \cite{Blitz2006} themselves formulated an alternative model for the computation of SFRs in galaxies, which seems more fundamental than the Schmidt--Kennicutt law. Applying both models for star formation to six molecule-rich galaxies in the local Universe, they showed that their new pressure-based law predicts SFRs nearly identical to the ones predicted by the Schmidt--Kennicutt law. Therefore, our choice to divide cold hydrogen in \ha~and \hm~according to pressure is indeed consistent with the prescription for SFRs used by \cite{DeLucia2007} and \cite{Croton2006}.

\subsection{Accuracy and limitations at $z=0$}\label{section_discussion_local}

A first limitation of our simulation comes from the assumption that the surface densities of \ha~and \hm~are axially symmetric (no spiral structures, no central bars, no warps, no satellite structures). In general, our model describes all galaxies as regular galaxies -- as do all semi-analytic models for the Millennium Simulation. Hence, the simulation results cannot be used to predict the \ha- and \hm-properties of irregular galaxies.

While our models allowed us to reproduce the observed relation between $\mha$ and $\rha$ remarkably well for various spiral galaxies (e.g.~Fig.~\ref{fig_mhi_rhi}), it tends to underestimate the size of \ha-distributions in elliptical galaxies. For example observations by \citet{Morganti1997} show that 7 nearby E- and S0-type galaxies all have very complex \ha-distributions, often reaching far beyond the corresponding radius of a mass-equivalent disk galaxies. The patchy \ha-distributions found around elliptical galaxies are probably due to mergers and tidal interactions, which could not be modeled in any of the semi-analytic schemes for the Millennium Simulation.

Another limitation arises from neglecting stellar bulges as an additional source of disk-pressure in Eq.~(\ref{eqpr}) \citep{Elmegreen1989}. Especially the heavier bulges of early-type spiral galaxies could introduce a positive correction of the central pressure and hence increase the molecular fraction, thus leading to very sharp \hm-peaks in the galaxy centers, such as observed, for example, in the SBb-type spiral galaxy NGC 3351 \citep{Leroy2008}. Our model for the \hm-surface density of Eq.~(\ref{eqsigmahm}) fails at predicting such sharp peaks, although the predicted total \ha- and \hm-mass and the corresponding radii and line profiles are not significantly affected by this effect.

\subsection{Accuracy and limitations at $z>0$}\label{section_discussion_evolution}

Additional limitations are likely to occur at higher redshifts, where our models make a number of assumptions based on low-redshift observations. Furthermore, the underlying DeLucia-catalog itself may suffer from inaccuracies at high redshift, but we shall restrict this discussion to possible issues associated with the models in this paper.

Regarding the subdivision of hydrogen into atomic and molecular material (Section \ref{section_masses}), our most critical assumption is the treatment of all galactic disks as regular exponential structures in hydro-gravitational equilibrium. This model is very likely to deviate more from the reality at high redshift, where galaxies were generally less virialized and mergers were much more frequent \citep{DeRavel2008}. Less virialized disks are thicker, which would decrease the average pressure and fraction of molecules compared to our model. Yet, galaxy mergers counteract this effect by creating complex shapes with locally increased pressures, where \hm~forms more efficiently, giving rise to merger-driven starbursts. Therefore, it is unclear whether the assumption of regular disks tends to underestimate or overestimate the \hm/\ha-ratios.

Another critical assumption is the high-redshift validity of the local relation between the \hm/\ha-ratio $\f$ and the external gas pressure $P$ (Eq.~\ref{eqblitz}). This relation is not a fundamental thermodynamic relation, but represents the effective relation between the average \hm/\ha-ratio and $P$, resulting from complex physical processes like cloud formation, \hm-formation on metallic grains, and \hm-destruction by the photodissociative radiation field of stars and supernovae. Therefore, the $\f-P$ relation could be subjected to a cosmic evolution resulting from the cosmic evolution of the cold gas metallicity or the cosmic evolution of the photodissociative radiation field. However, the metallicity evolution is likely to be problem only at the highest redshifts ($z\gtrsim10$). Observations in the local Universe show that spiral galaxies with metallicities differing by a factor $5$ fall on the same $\f-P$ relation \citep{Blitz2006}. Yet, the average cold gas metallicity of the galaxies in the DeLucia-catalog is only a factor $1.9$ ($3.6$) smaller at $z=5$ ($z=10$) than in the local Universe. These predictions are consistent with observational evidence from the Sloan Digital Sky Survey (SDSS) that stellar metallicities were at most a factor 1.5--2 smaller at $z\approx3$ than today \citep{Panter2008}. The effect of the cosmic evolution of the photodissociative radiation field on the $\f-P$ relation is difficult to assess. \citet{Blitz2006} argued that the ISM pressure and the radiation field both correlate with the surface density of stars and gas, and therefore the radiation field is correlated to pressure. This is supported by observations in the local Universe showing that the $\f-P$ relation holds true for dwarf galaxies and spiral galaxies spanning almost three orders of magnitude in SFR. For those reasons, the $\f-P$ relation could indeed extend surprisingly well to high redshifts.

In the expression for the disk-pressure in Eq.~(\ref{eqpr}) \citep{Elmegreen1989}, we assumed a constant average velocity dispersion ratio $\gs$. Observations suggest that $\vc/\vg$ decreases significantly with redshift \citep{ForsterSchreiber2006,Genzel2008}, and therefore $\gs$ perhaps increases. This would lead to even higher \hm/\ha-ratios than predicted by our model. However, according to Eq.~(\ref{eqfc}) this is likely to be a problem only for galaxies with $\ms>\mg$, while most galaxies in the simulation at $z>2$ are indeed gas dominated.

Regarding cold gas geometries and velocity profiles, the most important limitation of our model again arises from the simplistic treatment of galactic disks as virialized exponential structures. Very young galaxies ($\lesssim10^8\,\rm yrs$) or galaxies undergoing a merger do not conform with this model, and therefore the predicted velocity profiles may be unreal and the disk radii may be meaningless. This is not just a limitation of the simulation, but it reveals a principal difficulty to describe galaxy populations dominated by very young or merging objects with quantities such as $\rha$ or $\whafifty$, which are common and useful for isolated systems in the local Universe.

The radio line widths predicted by our model (Section \ref{section_radio_lines}) may be underestimated at high redshift, due to the assumption of a constant random velocity dispersion $\vg$. \citet{ForsterSchreiber2006} found $\vc/\vg\approx2-4$ for 14 UV-selected galaxies at redshift $z\approx2$. This result suggests that radio lines at $z\approx2$ should be about 20\%-30\% broader than predicted by our model, and therefore the evolution of the TFRs could be slightly stronger than shown in Fig.~\ref{fig_alltfr_evolution}.

In summary, the \ha- and \hm-properties predicted for galaxies at high redshift are generally uncertain, even though no significant, systematic trend of the model-errors could be identified. Perhaps the largest deviations from the real Universe occur for very young galaxies or merging objects, while isolated field galaxies, typically late-type spiral systems, might be well described by the model at all redshifts.

In Section~\ref{section_luminosities}, we ascribed CO(1--0)-luminosities to the \hm-masses using the metallicity dependent $X$-factor of Eq.~(\ref{eqxsim}). This model neglects several important aspects: (i) the projected overlap of molecular clouds, which is negligible in the local Universe, may become significant at high redshifts, where galaxies are denser and richer in molecules; (ii) the temperature of the cosmic microwave background (CMB) increases with redshift, hence changing the level population of the CO-molecule \citep{Silk1997,Combes1999}; (iii) the CMB presents a background against which CO-sources are detect; (iv) the molecular material in the very dense galaxies, such as Ultra Luminous Infrared Galaxies, may be distributed smoothly rather than in clouds and clumps \citep{Downes1993}; (v) the higher SFRs in early galaxies probably led to higher gas temperatures, hence changing the CO-level population\footnote{This list is not exhaustive, see \cite{Maloney1988,Wall2007} for an overview of the physical complexity behind the $X$-factor.}. \cite{Combes1999} presented a simplistic simulation of the cosmic evolution of $X$, taking the cosmic evolution of metallicity and points (i) and (ii) into account. They found that for an \hm-rich disk galaxy $\langle X\rangle$ increases by a factor 1.8 from redshift $z=0$ to $z=5$. This value approximately matches the average increase of $X$ by a factor 2 predicted by our simulation using the purely metallicity-based model of Eq.~(\ref{eqxsim}). This indicates that the effects of (i) and (ii) approximately balance each other. If a more elaborate model for $X$ becomes available, the latter can be directly applied to correct our CO-predictions. In fact, the $X$-factor only affects the CO(1--0)-luminosity $\lco$ calculated via Eq.~(\ref{eq_mhm}), but has no effect on the line properties considered in this paper, namely the line widths $\wcopeak$, $\wcofifty$, $\wcotwenty$ and the normalized luminosity densities $\lcocenter$, $\lcomax$.

\section{Conclusion}\label{section_conclusion}

In this paper, we presented the first attempt to incorporate detailed cold gas properties in a semi-analytic simulation of galaxies in a large cosmological volume. To this end, we introduced a series of physical prescriptions to evaluate relevant properties of \ha~and \hm~in simulated model-galaxies.

When applied to the DeLucia-galaxy catalog for the Millennium Simulation, our recipes introduce only one free parameter in addition to the 9 free parameters of the semi-analytic model of the DeLucia-catalog (see Table 1 in \citealp{Croton2006}). This additional parameter, i.e.~the cold gas correction factor $\correction$ (Section \ref{section_background}), was tuned to the cosmic space density of cold gas in the local Universe. The additional parameter $\xi$, describing the transfer of angular momentum from the halo to the disk (Section \ref{section_appl}), is not a free parameter for the hydrogen simulation, since it is fixed by the baryon mass--scale radius relation. In fact, we deliberately did not adjust $\xi$ to match the observed \ha-mass--\ha-radius relation of \cite{Verheijen2001}, in order to check the reliability of our models against this observation.

Based on the DeLucia-catalog, we produced a virtual catalog of $\sim3\cdot10^7$ per redshift-snapshot with various cold gas properties. This catalog represents an extension of the DeLucia-catalog, and it can be used to investigate a broad variety of questions related to \ha, \hm, CO and their cosmic evolution. The results presented in this paper have been restricted to some important examples, most of which could be compared directly to available observations and hence constitute key results for the verification of our simulation:

\begin{enumerate}
  \item Based on a pressure-model for the molecular content of cold gas, the simulation simultaneously reproduces the \ha-MF and the \hm-MF (resp.~the CO-LF) observed in the local Universe within the measurement uncertainties (Fig.~\ref{fig_local_mf1}).
  \item The simulated \ha-MFs for spiral and elliptical galaxies considered individually also match the observations for simulated galaxies with well-defined galaxy types (Fig.~\ref{fig_local_mf_by_type}).
  \item The simulated \ha-radii, imply a mass--radius relation for \ha~that matches the empirical counterpart (Fig.~\ref{fig_mhi_rhi}), thus confirming that the relation between $\mha$ and $\rha$ is such that the average \ha-density inside $\rha$ is $3.8\,\msun\rm\,pc^{-2}$ for all galaxies in the local Universe, although this value sensibly depends on the definition of $\rha$.
  \item The simulation predicts that the mass--radius relations for \ha~and \hm~are similar in the local Universe, but that their high-redshift evolution is completely different (Fig.~\ref{fig_radius_evolution}).
  \item The simulated widths of the \ha-radio emission lines of spiral galaxies are consistent with the empirical HI-TFR derived from the HIPASS spiral galaxies (Fig.~\ref{fig_alltfr_local}a); and the simulation predicts that there is no analog HI-TFR for elliptical galaxies.
  \item The simulated the stellar mass-TFR and the baryonic TFR reveal good agreement with the empirical TFRs for both spiral and elliptical galaxies in the simulation (Figs.~\ref{fig_alltfr_local}b, c).
  \item These TFRs are observable manifestations of a more fundamental relation between circular velocity and total dynamical mass, as suggested by the small scatter in the relation between $\whatwenty$ and $\mvir$ (Fig.~\ref{fig_alltfr_local}d).
  \item At higher redshift, the simulation predicts that the above TFRs remain valid (except for \ha~at $z\approx10$), but that their scatter increases and their zero-point is shifted towards higher velocities at fixed mass -- a fundamental prediction of hierarchical growth (Fig.~\ref{fig_alltfr_evolution}).
\end{enumerate}

The good match between simulation and observation regarding gas masses, disk sizes, and velocity profiles supports the models and recipes established in this paper. It also validates the semi-analytic recipes used by \cite{DeLucia2007} and supports the Millennium Simulation \citep{Springel2005} as a whole.

In forthcoming investigations the presented extension of the DeLucia-catalog towards cold gas properties could be used to investigate more elaborate questions. For example, what is the bias of the cosmic structure, for example of the power spectrum, revealed in \ha-surveys or CO-surveys compared to the underlying dark matter structure? How many \ha-sources can we expect to detect in future experiments performed by the SKA? Or how does the global \hm/\ha-ratio evolve with redshift and how does it relate to the observed evolution of the SFR density?

\acknowledgments
This effort/activity is supported by the European Community Framework Programme 6, Square Kilometre Array Design Studies (SKADS), contract no 011938. The Millennium Simulation databases and the web application providing online access to them were constructed as part of the activities of the German Astrophysical Virtual Observatory. D.~O.~thanks Gerard Lemson for his help in accessing the simulation data, as well as Erwin de Blok, Scott Kay, Raul Angulo, Carlton Baugh, and Carlos Frenk for fruitful discussions. Finally, we thank the anonymous referee for the helpful suggestions.


\end{document}